\documentclass[aps, prx, superscriptaddress, twocolumn, nofootinbib]{revtex4-2}
\usepackage[colorlinks, linkcolor=red, anchorcolor=green, citecolor=blue]{hyperref}
\usepackage{amsmath}
\usepackage{amssymb}
\usepackage{graphicx}
\usepackage{color}
\usepackage{braket}
\usepackage{orcidlink}
\usepackage{comment}

\graphicspath{{figures}}

\begin{document}

\title{Quantum thermalization of Quark-Gluon Plasma}

\author{Shile Chen}\email{csl2023@tsinghua.edu.cn}
\author{Shuzhe Shi}\email{shuzhe-shi@tsinghua.edu.cn}
\affiliation{Department of Physics, Tsinghua University, Beijing 100084, China}
\author{Li Yan}\email{cliyan@fudan.edu.cn}
\affiliation{Institute of Modern Physics, Fudan University, Shanghai 200433, China}

\begin{abstract}
The thermalization of quark gluon plasma created in relativistic heavy-ion collisions is a crucial theoretical question in understanding the onset of hydrodynamics, and in a broad sense, a key step to the exploration of thermalization in isolated quantum systems.
Addressing this problem theoretically, in a first principle manner, requires a real-time, non-perturbative method. To this end, we carry out a fully quantum simulation on a classical hardware, of a massive Schwinger model, which well mimics QCD as it shares the important properties such as confinement and chiral symmetry breaking. We focus on the real-time evolution of the Wigner function, namely, the two-point correlation function, which approximates quark momentum distribution. 
In the context of the eigenstate thermalization hypothesis and the evolution of entropy, our solution reveals the emergence of quantum thermalization in quark-gluon plasma with a strong coupling constant, while thermalization fails progressively as a consequence of the gradually increased significance of quantum many-body scar states in a more weakly coupled system. More importantly, we observe the non-trivial role of the topological vacuum in thermalization, as the thermalization properties differ dramatically in the parity-even and parity-odd components of the Wigner function.
\end{abstract}

\maketitle

\section{Introduction}
Quark-gluon plasma, a novel state of matter associated with quantum chromodynamics (QCD), is the focus of high energy nuclear physics. Through a large amount of high-precision meansurements at Relativistic Heavy-Ion Collider at the Brookhaven National Laboratory and the Large Hadron Collider at CERN, quark-gluon plasma has been identified as a perfect fluid with extremely small dissipations~\cite{Shuryak:2014zxa}. 

Fluidity of quark-gluon plasma requires thermalization, at least locally, which is challenging to understand theoretically. 
At a classical level, expansion of quark-gluon plasma can be dynamically characterized by viscous hydrodynamics, even far from local equilibrium, from which a universal hydrodynamic behavior emerges at a short time scale, 
a phenomena known as the hydrodynamic attractor~\cite{Heller:2015dha, Heller:2013fn, Romatschke:2017vte, Romatschke:2017acs, Blaizot:2017ucy, Spalinski:2017mel, Ambrus:2021sjg, Kurkela:2019set, Almaalol:2020rnu, Blaizot:2019scw, Blaizot:2020gql, Blaizot:2021cdv, Chen:2024pez}.  At a semi-classical level, where parton degrees of freedom dominate, thermalization of quark-gluon plasma can be achieved via scatterings among quarks and gluons, with respect to the Boltzmann equation description~\cite{Baier:2000sb,Kurkela:2015qoa,Blaizot:2011xf, Blaizot:2013lga, Berges:2012us, Berges:2014bba, Blaizot:2014jna, Xu:2014ega}. Albeit with a weak coupling, Boltzmann equation gives rise to analogous attractor behavior as in hydrodynamics, which as well supports the swift onset of hydrodynamics in quark-gluon plasma after initial nucleus-nucleus collisions~\cite{Martinez:2010sc, Martinez:2012tu, Ryblewski:2012rr, Strickland:2017kux, Strickland:2018ayk, Romatschke:2017ejr, Berges:2020fwq}. At a quantum mechanical level~\cite{Gelis:2024iar}, however, thermalization of quark-gluon plasma corresponds to a rather fundamental question that remains unresolved in general: 
in a high-energy collision process preserving unitarity, how a \textit{pure state} formed by two incident nuclei evolves and generates a highly excited and \textit{isolated many-body quantum system} that thermalizes rapidly?
Some recent reviews on the out-of-equilibrium hydrodynamics and thermalization of quark-gluon plasma can be found in Refs.~\cite{Romatschke:2017ejr, Alqahtani:2017mhy, Florkowski:2017olj, Berges:2020fwq,Shen:2020mgh}.

Classical thermalization can be understood through ergodicity, where the phase-space trajectory of a system fills an isoenergetic hypersurface. This ensures that, in the long-time limit, the system's evolution is effectively equivalent to a statistical ensemble average. In the quantum realm, where evolution is governed by unitary operators preserving time-reversal symmetry, ergodicity is expected to manifest in the Hilbert space---or at least within a small subset corresponding to a given energy. The concept of quantum ergodicity is closely related to the Eigenstate Thermalization Hypothesis (ETH)~\cite{Deutsch:1991msp, Srednicki:1994mfb, Rigol:2007juv}, which has been discussed for a variety of interacting quantum systems, including condensed matter~\cite{Rigol:2009zz, Biroli:2010xhz, Kaufman_2016} and cold atoms~\cite{Rigol:2007juv}. ETH states that for the isolated non-integrable quantum systems and few-body operators, the finitely excited energy eigenstates can be treated thermal, in the way that the operator expectation with respect to energy eigenstates approximates a microcanonical ensemble average, up to corrections that are suppressed with respect to the dimension of the Hilbert space.  Given ETH, quantum thermalization is understood as a direct consequence of the equivalence between the long-time average of the operator expectation and thermal ensemble average, regardless of initial conditions.

Although ETH successfully bridges the gap between the isolated many-body quantum system and thermal statistical ensemble, 
violation of the ETH condition, and correspondingly the absence of thermalization in an isolated quantum system, cannot be generically determined a prior. Obviously, ETH does not apply in general to all possible observables,
and it has also been noticed that ETH may be valid only for a fraction of the energy eigenstates, known as the weak ETH scenario~\cite{Biroli:2010xhz, mori2016weakeigenstatethermalizationlarge}. 
Beyond the apparent violation of ETH expected in integrable systems, ETH is also known to break down in the presence of quantum many-body localization (QMBL)~\cite{Anderson:1958vr}, both of the effects prevent the ergodicity of local quantum states. In contrast to the thermal states assumption in ETH, there could also be athermal eigenstates generically generated that break weakly the quantum ergodicity, such as the quantum many-body scar (QMBS)~\cite{Bernien_2017, Schecter:2019oej, Su:2022glk}.
We refer the readers to the following references for recent reviews of ETH~\cite{Mori:2017qhg, DAlessio:2015qtq}, QMBL~\cite{Nandkishore:2014kca, Ueda:2020ehs}, and QMBS~\cite{Moudgalya:2021xlu, Serbyn:2020wys}.

As an isolated many-body quantum system, quark-gluon plasma created in heavy-ion collisions is the highly excited state of QCD matter. Given further the fact that QCD is non-linear and non-integrable, one expects ETH generically applicable to the quantum thermalization of quark-gluon plasma.
In principle, verification of ETH in the context of quark-gluon plasma requires the diagonalization of QCD Hamiltonian, which however is notoriously difficult owing to not only the complexity in QCD dynamics, but also the limitation of computational resources. 
As a proxy of QCD, in the present work we focus instead on the Schwinger model~\cite{Schwinger:1962tp},
which is also known as the quantum electrodynamics in 1+1 dimensions. 
Schwinger model shares many essential features of QCD, in particular the quark confinement and chiral symmetry breaking in vacuum~\cite{Lowenstein:1971fc, Jayewardena:1988td, Smilga:1992hx, Adam:1993fc, Adam:1997wt, Coleman:1976uz, Adam:1995us, Adam:1996np, Adam:1996qk}. In the massless limit, the Schwinger model is solvable because it is equivalent to a non-interacting bosonic theory~\cite{Schwinger:1962tp}, while the integrability is broken by a finite fermion mass. 
Recently, there have been remarkable progresses in
the study of the massive Schwinger model on a lattice, using quantum simulations (see~\cite{Bauer:2022hpo, Bauer:2023qgm} for recent reviews) on quantum~\cite{Klco:2018kyo, Farrell:2023fgd, Farrell:2024fit} or classical~\cite{Zache:2018cqq, Ikeda:2020agk, Kharzeev:2020kgc, deJong:2021wsd, Xie:2022jgj, Belyansky:2023rgh, Florio:2023dke, Florio:2023mzk, Barata:2023jgd, Ikeda:2023zil, Ikeda:2023vfk, Lee:2023urk, Florio:2024aix, Ghim:2024pxe} devices,
as well as other field theories in general (see e.g.,~\cite{Li:2021kcs, Li:2022lyt, Li:2023kex, Czajka:2021yll, Carena:2022kpg, Yao:2023pht, Ebner:2023ixq, Ikeda:2024rzv, Hayata:2023puo, Hayata:2023pkw, Hidaka:2024zkd, Wu:2024adk, Carena:2024dzu, Qian:2024gph}). Particularly, ETH of plaquette and electric energy operators~\cite{Yao:2023pht, Ebner:2023ixq} and scar states~\cite{Hayata:2023puo} are found in $\mathrm{SU}(2)$ gauge theory. We refer the readers to~\cite{Bauer:2022hpo, Bauer:2023qgm} for 
recent reviews of quantum simulation of high energy physics.
Therefore, a model analysis of the quantum thermalization of a one-dimensional quark-gluon plasma is possible.

It is worth emphasizing that the realization of a high-energy field theory on a lattice not only renders the theory solvable within a specified energy scale, but also connects high-energy physics to practical lattice systems, such as spin chain. Given that such connections have been effectively used to gain insights into critical phenomena, universality, and phase transitions in quantum fields (cf. \cite{Rajagopal:1992qz}), it should come as no surprise that the Schwinger model on a lattice enables the study of the characteristics of quantum thermalization in quark-gluon plasma.

In the classical or semi-classical approach, a natural probe of thermalization is the single-particle phase-space distribution. Evolution of the distribution follows kinetic equation, with the well-known asymptotically equilibrium solution given by the Fermi--Dirac distribution (for fermions), Bose--Einstein distribution (for bosons), or Boltzmann distribution (for classical particles). In the long wavelength limit, moments of the phase-space distribution also determine fluid dynamical behavior in the system, which explains fundamentally the exhibition of hydrodynamic attractor in the kinetic theory solution~\cite{Blaizot:2019scw}.
For the quantum thermalization of quark-gluon plasma, as a quantum correspondence, we are motivated to consider the quark Wigner operators, which in principle, consists of the non-local two-body operators. More explicitly, the equal-time Wigner operator
\begin{align}
\label{eq:wignerf}
        \hat{W}_{\alpha\beta}(t,z,p) = \int \bar{\psi}_\alpha(z_+) U(z_+,z_-) \psi_\beta(z_-) \,e^{i\frac{py}{\hbar}}\,\mathrm{d}y\,,
\end{align}
can be obtained as the Wigner--Weyl transform of the gauge invariant two-point correlator of fermion field, $\psi$ and $\bar\psi$, at $z_\pm \equiv z \pm \frac{y}{2}$, with $U(z_+,z_-) \equiv e^{-i\frac{g}{\hbar}\int_{z_-}^{z_+} A(z) \mathrm{d}z}$ the gauge link ensuring gauge invariance.
Note that the momentum $p$ is introduced as the conjugate variable to spatial difference $y$. 
The expectation value of the Wigner operator gives rise to the Wigner function, which in the semi-classical limit, i.e., $\hbar\to 0$, reduces to quark phase-space distribution, upon an on-shell condition.

As a fermion in 1+1 dimensions, the quark Wigner operator $\hat{W}_{\alpha\beta}$ forms a $2\times2$ matrix (with indices $\alpha$ and $\beta=1,2$). The operator can be decomposed into different components by Dirac gamma matrices,
\begin{align}
     \hat{W} = \hat{w}_s - i\,\hat{w}_p\gamma^5 + \hat{w}_0 \gamma^0 + \hat{w}_1 \gamma^1\,.
\end{align}
Here, $\hat{w}_s$ and $\hat{w}_p$ are scalar and pseudo-scalar components, while $\hat{w}_0$ and $\hat{w}_1$ correspond to the vector charge and the axial vector charge components in 1+1 dimensions, respectively \footnote{
Particularly in 1+1 dimensions, $\hat{w}_0$ and $\hat{w}_1$ can be interpreted as well as the axial vector current and the vector current components.
}.
Given the unitarity condition $\hat W = \gamma^0 \hat W^\dagger \gamma^0$, the Wigner function $w_i$'s [$w_i\equiv \bra{\Psi(t)} \hat w_i\ket{\Psi(t)}$ is the expectation of Wigner operator measured at a time-evolving pure state, and $i\in\{s,p,0,1\}$] are real functions of $t$, $z$, and $p$ by construction.
Particularly, $w_0$ is the Fourier transformation of the two-point correlator ${\psi}^\dagger(z_+) \psi(z_-)$ and measures the difference between fermion and antifermion population, whereas $w_s$ corresponds to $\bar{\psi}(z_+) \psi(z_-)$ and measures the sum of population of fermions and antifermions. 
Meanwhile, the momentum integration of $w_0$ is the space-time dependent net charge density. 
Therefore, $(w_s+w_0)/2$ and $(w_s-w_0)/2$ respectively correspond to the quantum descriptions of the single-particle phase-space distribution functions of fermion and antifermion.
Analogously, $w_1$ is the Wigner transformation of ${\psi}^\dagger(z_+) \gamma^1 \psi(z_-)$ and corresponds to the phase-space distribution of chirality charge, and $w_p$ measures the difference between chirality charge carried by fermions and those carried by antifermions. 

As a preliminary study of the quantum thermalization of quark-gluon plasma, the current work focuses on the global equilibration in a finite 
and uniform system, hence we are allowed to average over the spatial coordinates, i.e., for system of size $L$, $w_i(t,p) \equiv \int w_i(t,z,p)\frac{dz}{L}$. 

Unlike solving the phase-space distribution in kinetic theory, evolution of the Wigner function is purely quantum mechanical. It does not rely on a weak-coupling assumption or a hierarchy that truncates the quantum equation of motion for the two-point correlation. Moreover, in the context of quantum simulation of the Schwinger model, the Wigner function is approachable irrespective of coupling strength. In fact, we are allowed to vary the coupling in our simulations, from weak to strong, and the nature of quantum thermalization in quark-gluon plasma differs dramatically. 
It is not our purpose in this current study to resolve the fundamental aspects of ETH, on whether and how it may or may not apply to a system, an operator, etc. It is our purpose to investigate, in the context of ETH, how QGP as a complex quantum many-body system emerges as a thermal system with finite system size and finite degrees of freedom (in terms of particle numbers).

The paper is organized as follows. After presenting the formulation of quantum simulation of the Schwinger model in Sect.~\ref{sec:qSchwinger}, we discuss the Wigner function thermalization in Sect.~\ref{sec:wignerTh}. With respect to ETH, the quantum thermalization of the Wigner function is re-interpreted in Sect.~\ref{sec:ETH}. In particular, we also explore the production of entropy, which is beyond the ETH approach (Sect.~\ref{sec:entropy}). Summary and discussion are given in Sect.~\ref{sec:summary}.
\begin{figure}[!hbtp]
    \centering
    \includegraphics[width=0.35\textwidth]{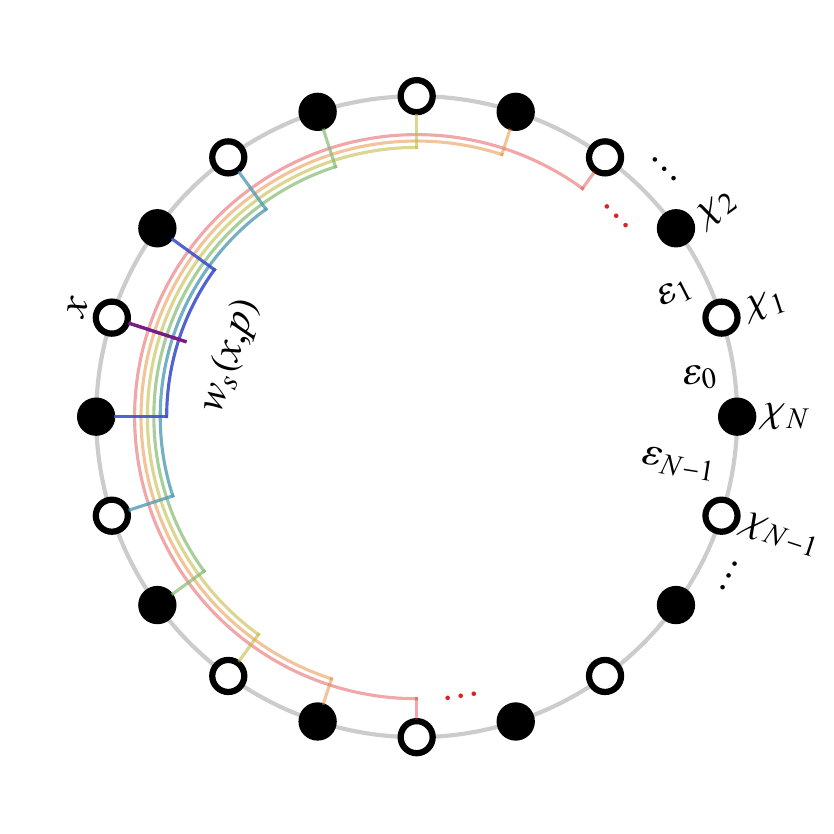}
    \caption{Illustration of the setup of a lattice Schwinger model with periodic boundary condition: filled and open symbols represent respectively the quark and anti-quark, which are linked by the electric field. Wigner operator defined with spacing is shown as well.
    \label{fig:illustration}}
\end{figure}

\section{Quantum simulation of the Schwinger model}
\label{sec:qSchwinger}

The Schwinger model characterizes interactions among fermion, anti-fermion, and an Abelian gauge field with a non-trivial topological vacuum solution characterized by a $\theta$-term, in 1+1 dimensions. 
In its action, 
\begin{equation}
S = \int \big( \bar{\psi}(\gamma^\mu (i\partial_\mu + g\, A_\mu)- m )\psi - \frac{F^{\mu\nu} F_{\mu\nu}}{4}  +\frac{g\theta}{4\pi}\epsilon^{\mu\nu}F_{\mu\nu}\big)\mathrm{d}^2x,
\end{equation} 
$\psi$ and $\bar\psi$ are fermion field operators, $F^{\mu\nu}=\partial^\mu A^\nu - \partial^\nu A^\mu$ gives the field strength, $m$ is the fermion (bare) mass and $g$ the coupling constant. Upon a chiral transformation that $\psi\to e^{i\frac{\theta}{2}\gamma^5} \psi$ and $\bar \psi\to\bar\psi  \,e^{i\frac{\theta}{2}\gamma^5}$, the $\theta$-term can be canceled due to anomaly, leading equivalently to the Hamiltonian
\begin{align}
    \hat{H} = \int 
    \Big( \bar{\psi}\big(\gamma^1 (-i\partial_z - g\, A_1) + m\,e^{i\theta\gamma^5} \big)\psi + \frac{1}{2}\mathcal{E}^2 \Big)\mathrm{d}z\,.
    \label{eq:Hamiltonian}
\end{align}
In Eq.~(\ref{eq:Hamiltonian}), $A_1$ is the gauge field potential which gives rise to the electric field $\mathcal{E}=\partial_0 A_1$. 
Throughout our discussion, to mimic the feature of quark-gluon plasma, we will consider these fermions as one-flavor quarks, and the coupling $g$ as the strong coupling constant.
Since what we discuss is purely quantum mechanical, for the remainder of the paper we do not distinguish quantum effects from the classical ones, and we employ the natural unit that $\hbar=c=k_B=1$.

In numerical simulations of quantum field theory, a finite volume must be used, with coordinate space discretized. Here, we consider a finite interval \( z \in [0, L] \) with periodic boundary conditions \([\psi(L) = \psi(0)\) and \(\mathcal{E}(L) = \mathcal{E}(0) = \mathcal{E}_\mathrm{bnd}]\) to maintain translational invariance, which is essential for defining a discrete Wigner function consistently.

The \( z \)-direction is discretized into a lattice with \( N \) grid points and spacing \( a = L/N \). In 1+1 dimensions, the Dirac matrices \( \gamma^\mu \) are \( 2 \times 2 \) matrices, and the field \( \psi \) has two degrees of freedom. On the lattice, we employ staggered fermions as introduced by Kogut and Susskind~\cite{Kogut:1974ag, Susskind:1976jm}, with \(\chi_{2n} = a^{\frac{1}{2}}\psi_{\uparrow}(2na)\) and \(\chi_{2n+1} = a^{\frac{1}{2}}\psi_{\downarrow}(2na + a)\). 
Using the Jordan-Wigner transformation~\cite{Jordan:1928wi}, the staggered fermions can be represented by Pauli matrices, ensuring the proper anticommutation relations. The Hamiltonian’s gauge invariance is maintained by implementing Gauss' law, \(\partial_z \mathcal{E} = g \, \bar{\psi} \gamma^0 \psi\), which constrains the gauge field operators at most lattice sites.
The only remaining independent degree of freedom is the electric field at the boundary. We introduce dimensionless operators, \(\varepsilon = g^{-1} \mathcal{E}_\mathrm{bnd}\) and \(U = e^{-i \, g \int_{0}^{L} A_1(z) \, \mathrm{d}z}\), to represent the electric and gauge fields, respectively. These operators satisfy the commutation relation \([\varepsilon, U] = -U\). For an approximate matrix representation of \(\varepsilon\) and \(U\), refer to~\cite{Shaw:2020udc}. An illustration of the staggered fermions and gauge fields on grids with period boundary condition is given in Figure~\ref{fig:illustration}.

In a system with $N$ staggered fermion grids and electric field truncation being $\Lambda$, we expand the quantum states by the Fock states $\ket{e;s_1,s_2,\cdots,s_N}$, in which $s_i \in \{0,1\}$ depending on whether or not the fermion/antifermion on the $i^\mathrm{th}$ site is occupied; and $e\in\{-\Lambda, \cdots, \Lambda-1, \Lambda\}$ is the truncated eigenvalues of the boundary electric field operator ($\varepsilon$). The Fock states therefore span a $(2\Lambda+1)2^N$ dimensional Hilbert space, and all operators are represented by $(2\Lambda+1)2^N$-by-$(2\Lambda+1)2^N$ dimensional matrices. Particularly, due to the Pauli exclusion principle, such a system has at most $N/2$ fermions and $N/2$ antifermions, whereas the number of gauge particles can be large.

Adopting the convention for gamma matrices in $1+1$ dimensions to be $\gamma^0 = Z$, $\gamma^1=iY$, and $\gamma^5 = \gamma^0 \gamma^1 = X$, with $X$, $Y$, and $Z$ being the Pauli matrices, the lattice Schwinger Hamiltonian reads, 
\begin{align}
\begin{split}
\hat{H}_{\mathrm{lat}}
=\;&
    \frac{i}{2}\sum_{n=1}^{N-1} 
    \Big((-1)^n m\sin\theta -\frac{1}{a}\Big)
    \Big(  \chi^\dagger_{n} \chi^{}_{n+1} - \chi^\dagger_{n+1}  \chi^{}_{n}\Big)
\\&
    +\frac{i}{2}\Big(m\sin\theta -\frac{1}{a}\Big)
    \Big(\chi^\dagger_{N} U^\dagger \chi^{}_{1} - \chi^\dagger_{1} U \chi^{}_{N}\Big)
\\&
    +\sum_{n=1}^{N}\Big( (-1)^n m\cos\theta\,\chi_n^\dagger \chi^{}_n
    + \frac{a\, g^2}{2}\varepsilon_n^2\Big)\,.
\end{split}
\label{eq:Hamiltonian_lat}
\end{align}
with the electric field given by
\begin{align}
    \varepsilon_n = \varepsilon + \sum_{m=1}^n \Big(\chi_m^\dagger \chi_m^{} - \frac{1-(-1)^m}{2}\Big)\,.
\end{align}
While the Hamiltonian~\eqref{eq:Hamiltonian_lat} is ready to be simulated on quantum devices, our real-time simulations require a large quantum circuit depth that exceeds the current capabilities. In this work, we simulate the evolution of quantum states on classical hardwares, i.e., solve the time-dependent state vector governed by the Hamiltonian matrix. A detailed derivation of the lattice Schwinger Hamiltonian, including its explicit gate representation, is provided in Appendix~\ref{sec:app:schwinger_model}.

In 1+1 dimensions, the coupling constant $g$ has the dimension of energy, just as the quark mass $m$. In a lattice theory, from the grid size one has $1/a$ and $1/(Na)$ respectively the ultraviolet (UV) and infrared (IR) cut-off of the energy scale. In this paper, we always scale the dimensionful quantities in units of $g$. 

Unless otherwise specified, we will focus on the case of $\theta=0$ in what follows.
The lattice Schwinger model~\eqref{eq:Hamiltonian_lat} describes a fermion system inside a staggered external field (terms $\propto (-1)^n m$) with hopping between adjacent fermion and antifermion (terms $\propto 1/2a$), while additionally a gauge field link ($\propto a g^2/2$) applies inducing confinement and vacuum chiral condensate. 
With respect to the Jordan--Wigner transformation, the lattice Schwinger Hamiltonian is dual to a 1-dimensional spin chain\footnote{Note that the surface gauge field with proper truncation is dual to a site with high spin.}, for which the properties of ETH have been studied quite extensively.

In principle, one would expect the mass term in the Hamiltonian suppresses local fluctuations of fermion/antifermion occupation from the bare vacuum, i.e., the N\'eel state. In addition, although the electric field energy formally contains all-to-all connections\footnote{One may find $\chi_m^\dagger \chi_m^{} \chi_n^\dagger \chi_n^{}$ with arbitrary $m$ and $n$ when expanding $\sum_n \varepsilon_n^2$.}, it disfavors configurations with electric field link that is strong in intensity or long in distance. Thus, the long-range correlations are suppressed by the electric field energy term. In contrast, the hopping terms, which are summation of creation/annihilation operators of nearest pairs, bring the ground state away from the trivial bare vacuum. It has been found, in the absence of the mass term, that a strong enough hopping (corresponding to sufficiently fine grids) is essential for breaking the localization induced by the electric field energy~\cite{Brenes:2017wzd}. 
Nonetheless, the effective electric localization is avoided in cases of finer lattice spacing, which corresponds to a higher energy scale that we are considering.
In our numerical simulation, unless specified, we take the largest lattice grid\footnote{The quantum states span a Hilbert space with dimension $\mathcal{D} = (2\Lambda+1)2^N = 5\times2^{20} \approx 5.2\times 10^{6}$. Noting that the total charge operator commutes with the Hamiltonian and Wigner operators, we focus on the neutral sector which reduces the dimension by a factor $\sim 5$. Thus, all operators are one-million-by-one-million dimensional matrices.} allowed by the computation resource, $N=20$, with spacing $a=1/(2g)$ and electric field cut-off $\Lambda=2$. They correspond to energy cut-off scales $\Lambda_\mathrm{UV} = 2g$ and $\Lambda_\mathrm{IR} = g/10$. 

It is interesting to notice that the system size of the lattice $Na = 10/g= 18 M_\mathrm{meson}^{-1}$
\footnote{
The massless Schwinger model in the continuum limit is exactly solvable, as it is dual to a free meson system with meson mass $M_\mathrm{meson}=g/\sqrt{\pi}$~\cite{Schwinger:1962tp}.},
is comparable with the typical system size in a heavy-ion collision, which is $\sim 10\, \mathrm{fm} = 15\,M_\pi^{-1}$ ($M_\pi = 135~\mathrm{MeV}/c^2$ is the neutral pion mass). Additionally, with such lattice spacing, thermalization has been observed in the entanglement spectrum in jet production~\cite{Florio:2023dke, Florio:2024aix} as well as entanglement entropy and chiral condensate under the presence of random external electric charges~\cite{Brenes:2017wzd}, and consistency with the continuum limit has been examined~\cite{Florio:2023dke, Florio:2024aix}.

We then explore various values of fermion mass between $0$ and $\Lambda_\mathrm{UV}=2g$ to study whether or how the system approaches thermal equilibrium. Note that although the massless Schwinger model in the continuum limit is exactly solvable, 
its discrete version is not integrable regardless of $m$. This can be understood intuitively from the perturbative calculation of mass renormalization, and it is found that a finite lattice effect of Schwinger model induces a correction $\delta m = a\,g^2/8$ in the bare mass~\cite{Dempsey:2022nys}. Thus, instead of being a massless fermion system, the $m=0$ case should be treated as a strongly coupled scenario that has large ratio of $g/m_\mathrm{eff}$. Correspondingly, while hitting the UV cut-off, the $m=2g$ limit can also be viewed as a weakly coupled scenario that $g/m=1/2$. In what follows, we refer to the $m=0$ and $m=2g$ settings as strong and weak coupling cases, respectively.

On a lattice, the momentum space is also discretized to maintain the periodic boundary condition. Thus, the Wigner operators ought to be defined at momentum grids, $\hat{w}_{i,k} \equiv \hat{w}_i(p=\frac{\pi}{N\,a} k)$ with $k$ being integer, and they are represented by 
\begin{align}
\begin{split}
\hat{w}_{s,k} =\;& \sum_{n=1}^{N}\sum_{m=0}^{N-1} 
    \frac{e^{i\frac{2m\pi k}{N}}}{N}
    (-1)^{n-m} \chi_{n+m}^\dagger \chi_{n-m}^{}\,,
\end{split}
    \label{eq:W_s}\\
\begin{split}
\hat{w}_{0,k} =\;& \sum_{n=1}^{N}\sum_{m=0}^{N-1}
    \frac{e^{i\frac{2m\pi k}{N}} }{N}
    \chi_{n+m}^\dagger \chi_{n-m}^{}\,,
\end{split}
    \label{eq:W_v}\\
\begin{split}
\hat{w}_{1,k} =\;& \sum_{n=1}^{N}\sum_{m=0}^{N-1}
    \frac{e^{i\frac{(2m+1) \pi k}{N}}}{N}
    \chi_{n+m}^\dagger \chi_{n-m-1}^{}\,,
\end{split}
    \label{eq:W_a}\\
\begin{split}
\hat{w}_{p,k} =\;& \sum_{n=1}^{N}\sum_{m=0}^{N-1}
    \frac{i\,e^{i\frac{(2m+1) \pi k}{N}}}{N}
    (-1)^{n+m} \chi_{n+m}^\dagger \chi_{n-m-1}^{}\,.
\end{split}
    \label{eq:W_p}
\end{align}
It is obvious that $\hat{w}_{i}(p+\frac{\pi}{a}) = \pm \hat{w}_{i}(p)$, with positive and negative signs taken for $i\in\{s,0\}$ and $\{p,1\}$, respectively.

\begin{figure}
    \centering
    \includegraphics[width=0.235\textwidth]{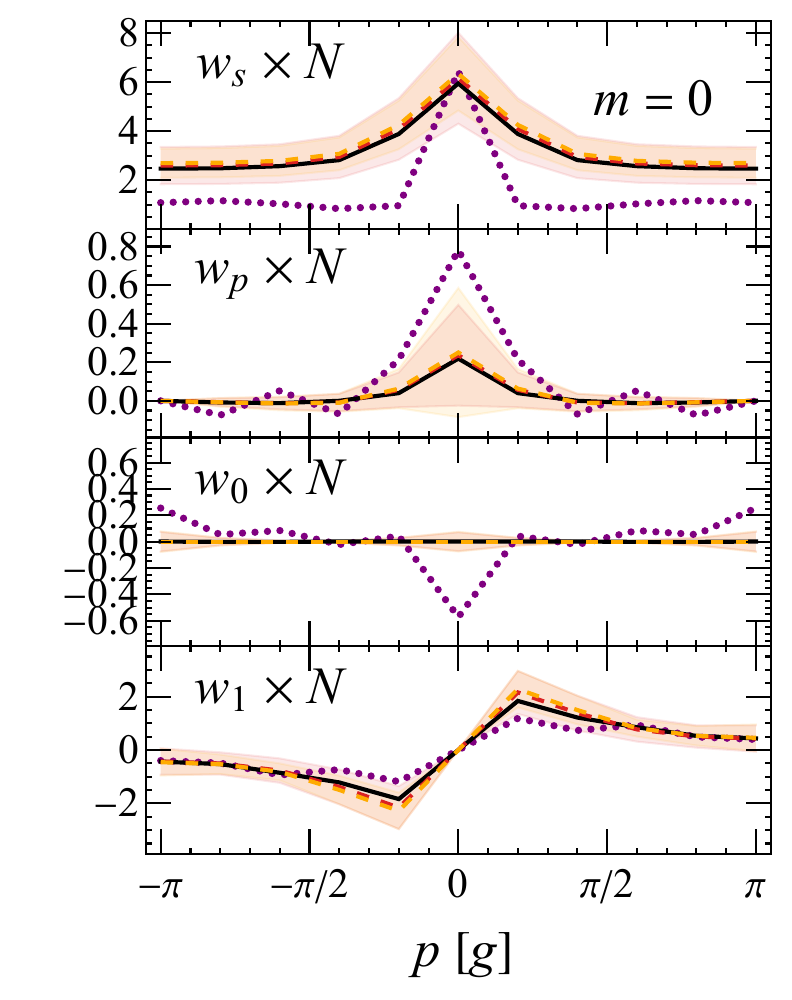}
    \includegraphics[width=0.235\textwidth]{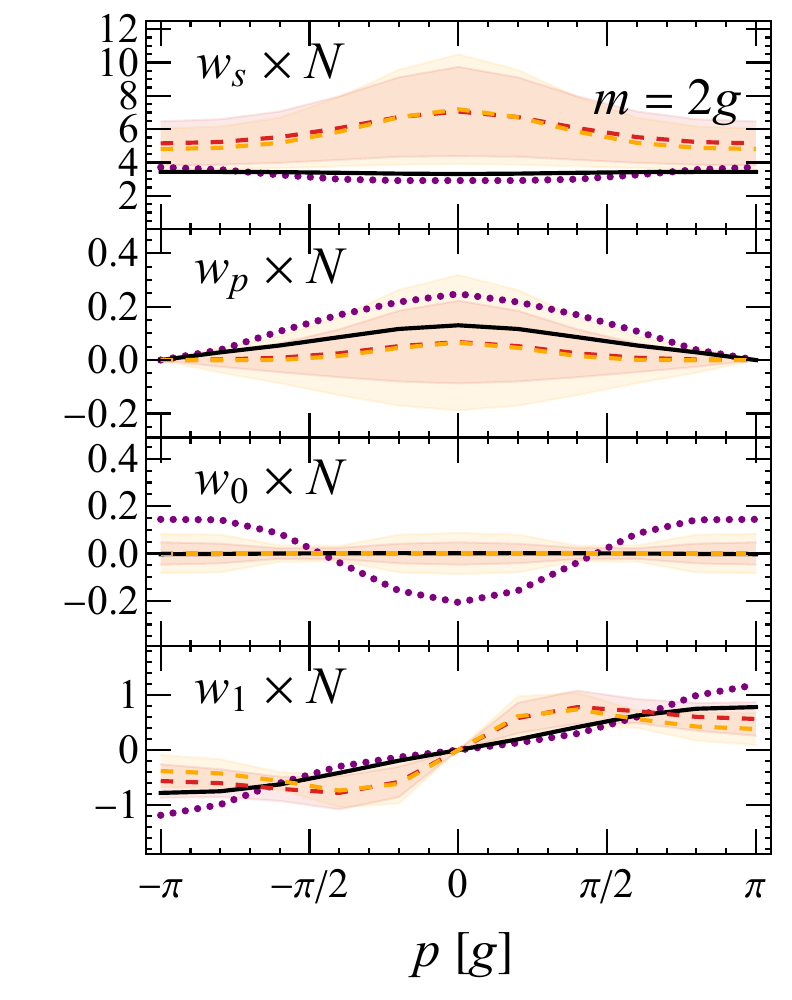}
    \caption{
    (Top to bottom) Momentum dependence of scalar, vector, pseudoscalar, and axialvector components of the Wigner function. 
    Purple dotted lines indicate initial value, and black solid lines show the long-time averages.
    Gold (red) lines correspond to the thermal expectation values using MCE (CE), with the bands are the corresponding thermal variation. 
    \label{fig:wigner}}
\end{figure}
\section{Wigner function thermalization}
\label{sec:wignerTh}

With an initial state $\ket{\Psi(0)}$, the quantum state at time $t$ is given by 
\begin{align}
    \ket{\Psi(t)} = e^{-i\hat{H}t}\ket{\Psi(0)}\,,\label{eq:schroedinger}
\end{align}
and the Wigner functions are the expectation values of the corresponding operators, $\langle \hat{w}_i(p)\rangle_t\equiv \bra{\Psi(t)} \hat{w}_i(p)\ket{\Psi(t)}$ .
We simulate the quantum state evolution using a classical hardware with details provided in the Appendix~\ref{sec:app:state}.
Eventually, quantum thermalization of the Wigner operators can be captured by a direct comparison between quantum expectations $\langle \hat{w}_i(p)\rangle_t$ in the long-time limit 
and the thermal ensemble averages described as follows.

For a quantum system with a specified energy, one generically adopts a microcanonical ensemble description for the energy eigenstates, with the microcanonical ensemble average given by state average within a finite energy shell,
\begin{align}
\langle \hat{w}_i(p) \rangle_{\rm MCE} = \frac{1}{\cal N} \sum_{E'\in[E-\frac{\Delta E}{2},E+\frac{\Delta E}{2}]} \bra{E'} \hat{w}_i(p)\ket{E'}\,,
\end{align}
where ${\cal N}$ is the number of states inside the energy shell,
 $E=\bra{\Psi(t)}\hat H\ket{\Psi(t)}$ and $(\Delta E)^2 = \bra{\Psi(t)}\hat H^2\ket{\Psi(t)} - E^2$ are respectively the mean and variance of energy\footnote{It is straightforward to see that $E$ and $\Delta E$ are determined by the initial state and do not evolve in time.}. Alternatively, if the energy eigenstates are approximately characterized by a canonical ensemble, one has the canonical ensemble average through the density matrix $\hat \rho(T) \propto \exp(-\hat H/T)$,
\begin{align}
\langle \hat{w}_i(p) \rangle_{\rm CE} = \frac{{\rm tr} [e^{-\hat H/T} \hat{w}_i(p)]}{{\rm tr}[e^{-\hat H/T}]}\,.
\end{align}  
Here, the temperature of the canonical ensemble can be solved according to $E = \bra{\Psi(t)} \hat H \ket{\Psi(t)} = \langle \hat H\rangle_{\rm CE}$.

Microcanonical and canonical ensemble averages are respectively represented by the gold and red lines in Fig.~\ref{fig:wigner}. Note that due to the neutrality condition imposed in the system, the ensemble average of the vector charge component $w_0$, which corresponds to the difference between fermion and antifermion, vanishes by construction. Consequently, the scalar component alone characterizes the population of quarks.
We also estimate the strength of thermal fluctuations via the thermal variance, shown in Fig.~\ref{fig:wigner} as the colored bands. 
For instance, instead of a static distribution in thermal equilibrium, the equilibrium expectation of the scalar component fluctuates around its mean within a thermal width $\delta_{s}^{\rm MCE}(p) = (\langle \hat{w}_s(p)^2\rangle_{\rm MCE} - \langle \hat{w}_s(p)\rangle_{\rm MCE}^2)^{1/2}$. 
Thermal variance formulated in terms of the canonical ensemble is similar.

Knowledge of the equilibrium distribution of the Wigner function allows one to avoid quantum typicality in the initial condition. In practice, with respect to energy $E$, although initial state can be chosen arbitrarily on a patch of a unit hypersurface, namely,
\begin{align}
\ket{\Psi(0)} = \sum_n c_n \ket{E_n}\;\leftrightarrow\;\bra{\Psi(0)}\hat H\ket{\Psi(0)}=E
\label{eq:state}
\end{align}
with $\sum_n |c_n|^2 =1$, we find that a certain fraction of the unit hypersurface is quantum typical. 
These typical states lead to thermalization of the Wigner function operators regardless of interaction and evolution, e.g., $\bra{\Psi(0)} \hat{w}_i(p)\ket{\Psi(0)}= \langle \hat{w}_i(p)\rangle_{\rm CE}$. 
We therefore purposely choose the initial condition of the quantum state corresponds to energy level occupations far away from thermal equilibrium, with a scheme detailed in Appendix~\ref{sec:app:initial}. 
One such initial state results in expectations of initial Wigner operators, as shown with the purple dotted lines in Fig.~\ref{fig:wigner}, where deviations from the thermal ensemble averages are obvious.

With respect to the chosen initial condition, by solving the state~\eqref{eq:schroedinger}, one obtains the evolution of the Wigner functions. 
For instance, as shown in the left panel of Fig.~\ref{fig:wigner}, in the strong coupling case, although with quantum fluctuations, $w_s(p)$ gradually accumulates at all momenta, reflecting the generation of quarks and anti-quarks from gauge field. For all the four components in the strong coupling case, we observe that the long-time averaged expectations (black-solid lines) are consistent with their corresponding thermal ensemble averages. On the other hand, in the weak coupling case where quark mass $m=2g$, the long-time averaged expectations of the scalar and the axial-vector charge components differ significantly from thermal ensemble averages, while within the thermal variance, $w_0$ and $w_p$ still thermalize.
We have tested the solutions with various initial conditions.

\begin{figure}
    \centering
    \includegraphics[width=0.5\textwidth]{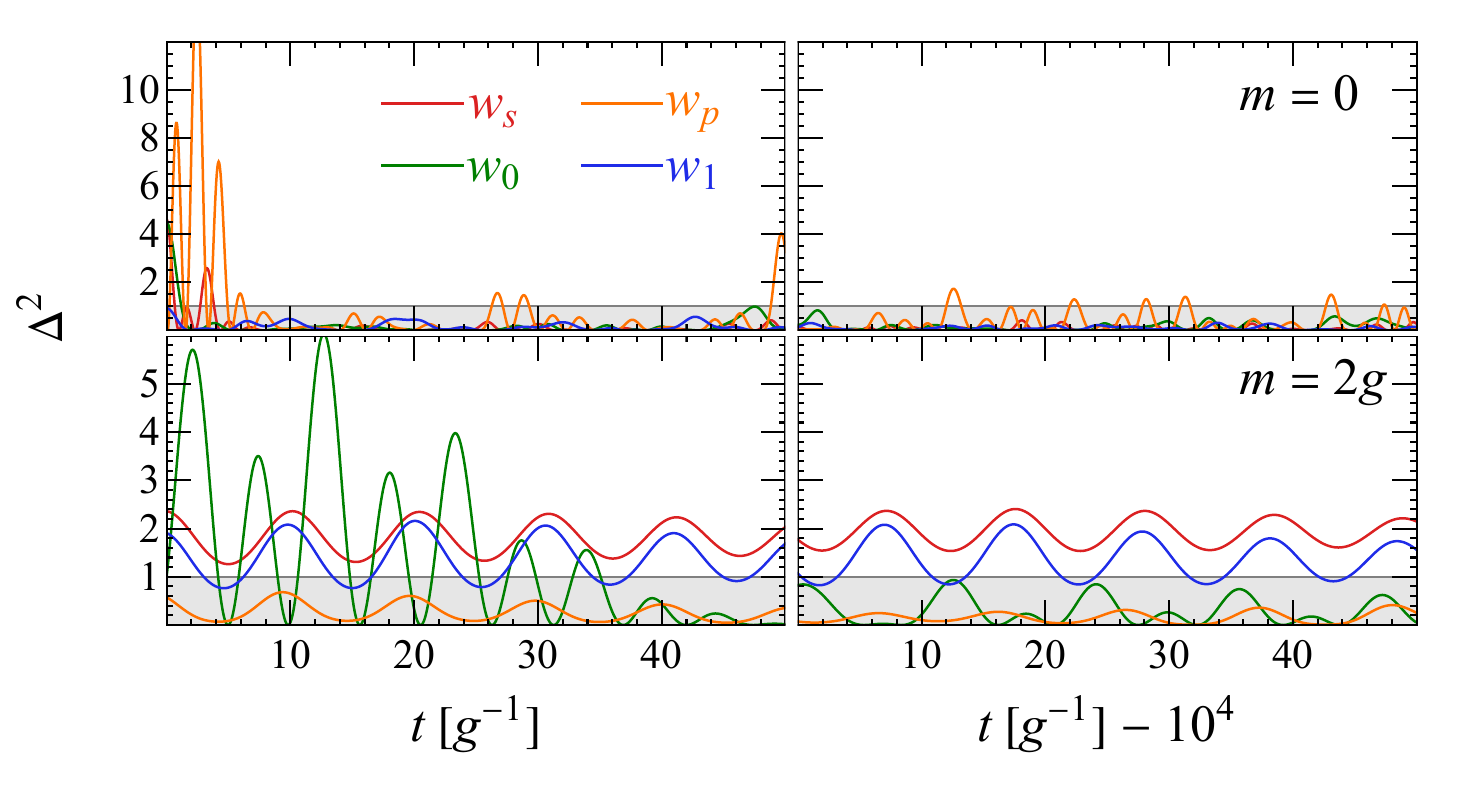}
    \caption{Time evolution of summed difference between the scalar(red), pseudoscalar(orange), vector(green), and axial vector(blue) components of the Wigner function and their corresponding thermal values. The top (bottom) panel is for $m/g=0$ ($2$). Left and right panels show early and late time evolution for ranges $t \in [0, 50]/g$ and $t \in [10000, 10050]/g$, respectively. A shaded band is added to indicate the thermal fluctuation region that $\Delta^2 \leq 1$.
    \label{fig:wigner_diff}}
\end{figure}

In order to achieve a time-dependent and quantitative measure of quantum thermalization, we further define the relative deviation,
\begin{align}
\Delta_i^2(p,t) \equiv \left(\frac{\langle \hat{w}_i(p)\rangle_t - \langle \hat{w}_i(p)\rangle_\mathrm{MCE}}{\delta_i^\mathrm{MCE}(p)}\right)^2\,,
\end{align}
where $i\in\{s,p,0,1\}$ and $\delta_i^{\rm MCE}$ is the thermal width introduced previously. 
This relative deviation is a non-negative quantity, and it vanishes only for absolute thermalization that Wigner function coincides with the thermal ensemble average. Thermal width in the denominator takes into account the effect of thermal fluctuations, such that thermalization can be recognized as long as $\Delta_i^2<1$.

In Fig.~\ref{fig:wigner_diff}, the evolution of $\Delta_i^2$'s at a specified momentum are shown, respectively, for the strong-coupling (top panels) and weak-coupling (lower panels) systems. 
For illustrative purposes, we choose the momentum separately for each curve, such that each of them has the most significant deviation from their corresponding thermal ensemble averages. In principle, Fig.~\ref{fig:wigner_diff}
conveys the message in line with Fig.~\ref{fig:wigner}, that all components of the Wigner function thermalize in the strong coupling system, while in a weak coupling system, quantum thermalization is achieved only in $w_p$ and $w_0$.  Additionally, time dependence allows one to explore the process of relaxation toward thermalization. 
For the component that thermalizes, the relative deviation starts from a large value and decay.  
After a relaxation time scale the relative deviation becomes generally smaller than unity, even quantum fluctuations occur occasionally and randomly bring them back away from zero. 
Note that as thermalization being approached, quantum fluctuations also decay significantly and eventually become comparable in magnitude to the thermal widths, indicating compatibility between quantum and thermal fluctuations. 
For a component that does not thermalize --- for instance, the scalar components $w_s$ in the weak coupling system, as shown in the lower panel of Fig.~\ref{fig:wigner_diff} --- the relative deviation and the amplitude of the quantum fluctuations do not decay in time, with $\Delta_s^2>1$. The evolution of the relative deviation in the long time limit is investigated as well, as we shift the time window with a delay $t\to t+10^4/g$. 

Our observations provide strong evidence for quantum thermalization in a strongly coupled quark-gluon system in 1+1 dimensions, as all components of the Wigner function at different momenta converge to the thermal distributions. In contrast, in a weakly coupled quark-gluon system, quantum thermalization can be realized only in the vector charge
and the pseudo-scalar components.

\begin{figure}
    \centering
    \includegraphics[width=0.235\textwidth]{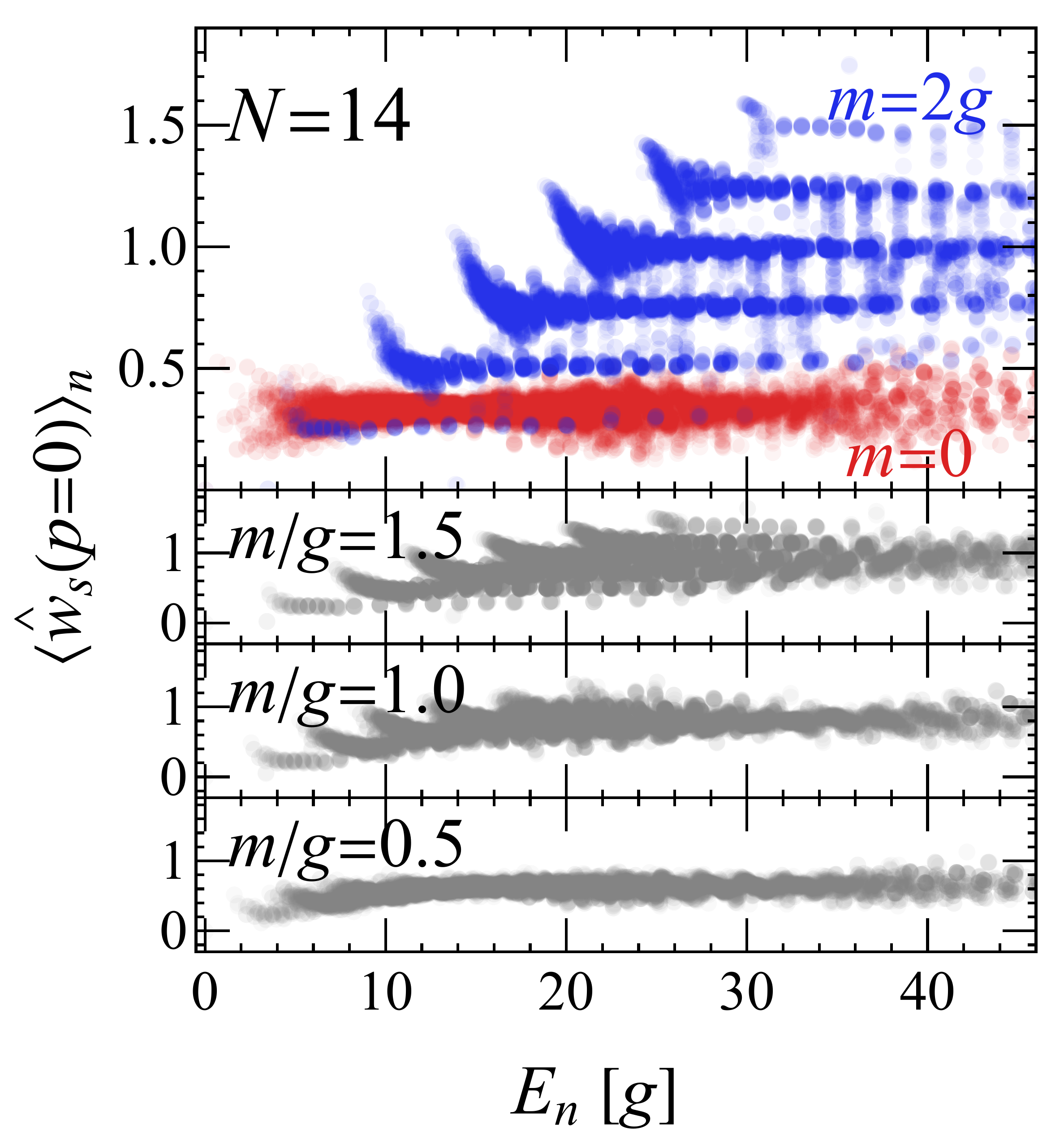}
    \includegraphics[width=0.235\textwidth]{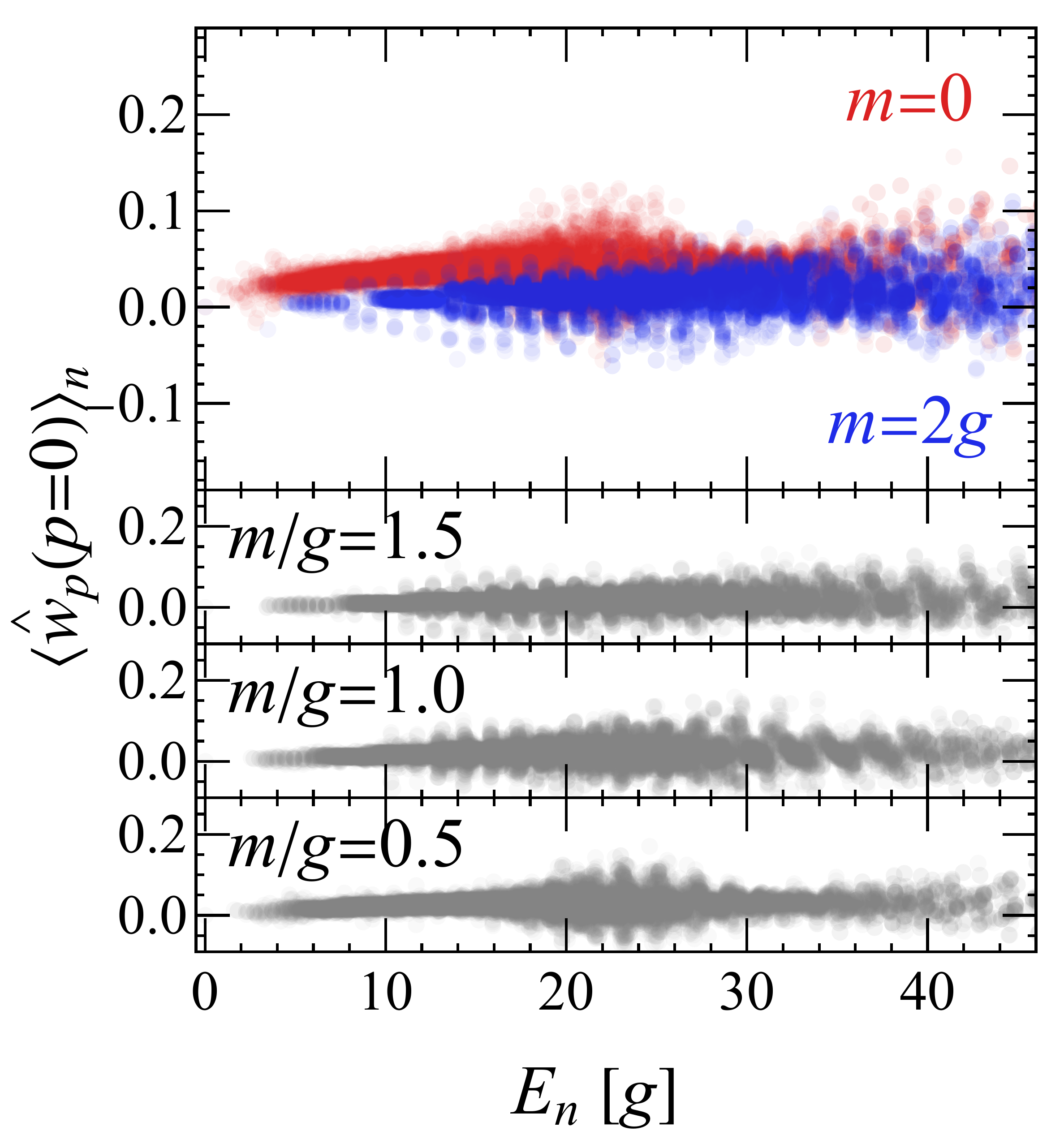}
    \caption{Expectation values of scalar Wigner function with $N=14$ and $m=0$ (red) or $m=2g$ (blue), with other values for mass are also shown in the lower panels. Wigner functions at different momentum points are qualitatively similar, and we show results at $p=0$ as representives. 
    \label{fig:eth}}
\end{figure}

\section{Eigenstate Thermalization Hypothesis}
\label{sec:ETH}

The thermalization of the Wigner function in the 1+1 dimensional quark-gluon system can be understood with respect to eigenstate thermalization hypothesis (ETH)~\cite{Deutsch:1991msp, Srednicki:1994mfb, Rigol:2007juv}. 
For the purpose of illustration, in this section we focus on the scalar component $w_s$ and the pseudo-scalar component $w_p$.

For a non-integrable quantum system, if ETH applies to an operator, denoted as $\hat O$, the matrix elements in the energy basis $O_{nm}\equiv \bra{E_n} \hat O\ket{E_m}$ can be expressed as, 
\begin{align}
    O_{nm} = O(\bar E)\delta_{nm} 
    + \Omega(\bar E,E_n-E_m) r_{nm}\,.
    \label{eq:eth}
\end{align}
The diagonal elements are dominated by a smooth function $O(\bar E)$ of $\bar E \equiv (E_n+E_m)/2$; while the off-diagonal elements, in addition to the random noise of order unity, $r_{nm}$'s,  are constrained by the function $\Omega(\bar E,E_n-E_m)\propto \exp(-S(\bar E)/2)$, which represents the suppression with respect to the density of microscopic states, which is also linked with the system's total entropy.
As a direct consequence of Eq.~(\ref{eq:eth}), given a quantum state, the expectation $\hat{O}$  is determined by the mean energy ($\langle E \rangle \equiv \sum_n p_n E_n$),
\begin{align}
    \langle \hat{O} \rangle \equiv \sum_n p_n O_{nn} = O(\langle E \rangle) \,(1 + \mathcal{O}(\exp(-S)).
\end{align}
Here, $p_n$ is the $n^\mathrm{th}$ diagonal of the density matrix, namely the occupation probability of the $n^\mathrm{th}$ energy level. 
It corresponds to the thermal weight [$p_n = e^{-\beta E_n}/Z(\beta)$] in a canonical ensemble, or the norm-squared of the expansion coefficient ($p_n = |c_n|^2$) for a pure state taking the form as Eq.~\eqref{eq:state}. Given that ETH is satisfied, thermalization takes place since the long-time average of the operator expectation $\lim_{t\to \infty}\langle \hat O\rangle_t = O(\langle E \rangle) \,(1 + \mathcal{O}(\exp(-S))$ approximates to the thermal value once the effective temperature is determined by matching the mean energy ($\langle E \rangle$). Furthermore, contributions from the off-diagonal terms are suppressed substantially by the inverse of density of states.

\vspace{3mm}
\textit{Distribution of Eigenstate spectrum.} 
We present in Fig.~\ref{fig:eth} the diagonal elements of the Wigner function $w_s$ and $w_p$ at momentum $p=0$, with the notation $\langle \hat{O} \rangle_n \equiv \bra{E_n} \hat{O} \ket{E_n}$. The off-diagonal elements of these Wigner functions are found essentially small (see Appendix~\ref{sec:app:eth}). 
At all other momenta, the Wigner function exhibits qualitatively the same, even quantitatively consistent behavior. In the current section, we perform the calculation for a smaller lattice grid with $N=14$ sites in order to check all the energy eigenstates. For the low excitation states that are achievable in larger lattice grids, we have checked that results are quantitatively consistent\footnote{Consistency can also be seen by the system size dependence examination performed in Appendix~\ref{sec:app:eth}.}.
In the left panel of Fig.~\ref{fig:eth}, in the case of $m=0$, the diagonal elements of $w_s$ are distributed within a narrow band, represented by a smooth function of energy, in accordance with the ETH condition. However, as the quark mass increases from zero, a splitting in the distribution of the diagonal elements is observed. In the case of an extremely weakly coupled system with $m=2g$, the diagonal elements are distributed into five distinct sub-bands, signaling a clear breakdown of the ETH condition. On the other hand, the diagonal elements of the pseudoscalar component of the Wigner function form a smooth band, irrespective of the quark mass, as shown in the right panel of Fig.~\ref{fig:eth}. 

With respect to the ETH conditions, all the properties of the matrix elements are consistent with our previous observation, that the thermalization of $w_s$ is dependent on the quark mass, while the pseudoscalar component of the Wigner function consistently thermalizes, regardless of the quark mass.

\begin{figure}[!hbtp]
    \centering
    \includegraphics[width=0.45\textwidth]{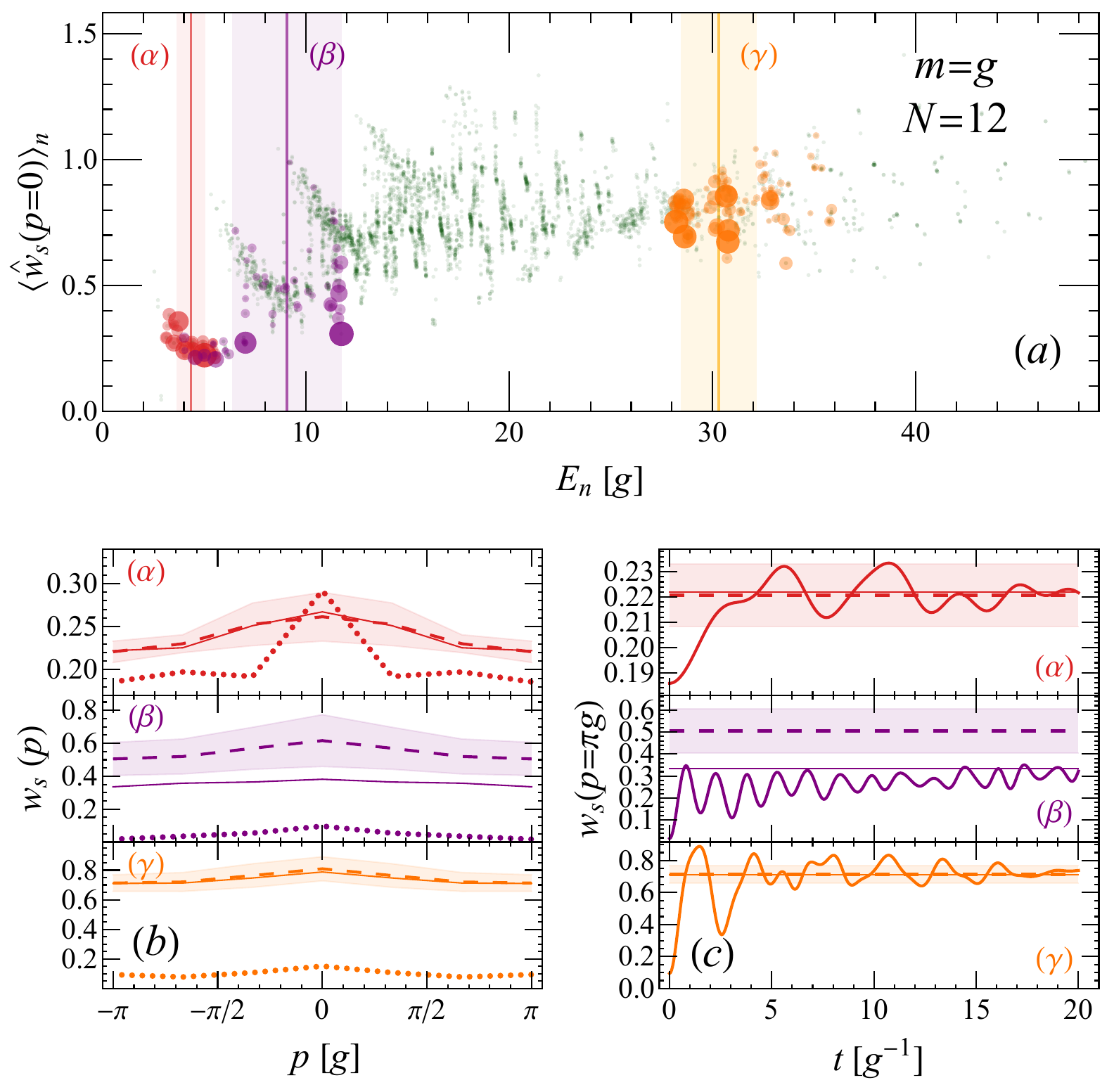}
    \caption{($a$) Same as \protect{Fig.~\ref{fig:eth}} but for $m=g$. The size of red, purple, and orange circles indicate the probability of the corresponding energy levels in the simulations labeled as ($\alpha$), ($\beta$), and ($\gamma$), respectively. Vertical lines(bands) indicate the mean values(standard derivations) of energy of the corresponding quantum state. ($b$) Momentum dependent scalar Wigner functions. Dotted, solid, and dashed lines and bands respectively represent the initial, long time averaged, and MCE thermal averaged values and fluctuations. ($c$) Time evolution of the scalar Wigner function with momentum $p=\pi g$. Thin horizontal solid lines represent the long-time-averaged values, and thick dashed lines and bands are respectively the MCE thermal averages and fluctuations. 
    \label{fig:eth_2}}
\end{figure}

\vspace{3mm}
\textit{Time evolution of the scalar component.} 
Let us now focus on the scalar component of the Wigner function.
Noticing that the splitting in the diagonal elements appears gradually in the low energy region, as quark mass increases, the violation of ETH condition could be only partial, depending on the system energy. This is the weak-ETH scenario in a non-integrable system~\cite{Biroli:2010xhz, mori2016weakeigenstatethermalizationlarge}.
The dependence of thermalization on the system energy can be best demonstrated for a intermediate mass value, such as $m=g$, as to some extent it is a pseudo-critical mass in the system.
At particularly low energies, only one branch of the split bands affects, such as the energy at line $(\alpha)$ in the top panel of Fig.~\ref{fig:eth_2} (panel (a)), the diagonal elements can still be regarded being distributed as a part of a smooth function, corresponding to microcanonical ensemble in a smaller Hilbert space.
Analogous argument applies as well, for sufficiently large energies where the sub-bands converge, such as the energy at line $(\gamma)$ in the top panel of Fig.~\ref{fig:eth_2}.  For both cases, one would still expect thermalization. 
However, thermalization of the scalar Wigner function should be absent for the energy at line $(\beta)$, given the apparent violation of the ETH that can be observed in the multiple sub-bands in the distribution of the diagonal elements contribute.

If one solves the system evolution for the 
scalar Wigner function at these energies, thermalization behavior regarding ETH can be verified, which are shown in the lower panels of Fig.~\ref{fig:eth_2}.  In the lower left panel (panel (b)), as energy varies, the long-time average of the scalar Wigner function is compared to the thermal ensemble average.  In the cases of $(\alpha)$ and $(\gamma)$, starting from an out-of-equilibrium state (dotted lines), the long-time averages of the scalar Wigner function (solid lines) converge correspondingly to their thermal ensemble averages (dashed lines with bands). In case $(\beta)$, contrastingly, thermalization is absent. The lower right panel (panel (c)) confirms the observation, with emphasis on time evolution. For both $(\alpha)$ and $(\gamma)$, $w_s(0)$ relaxes towards the thermal expectation. Note that since the relaxation time scales inversely to the dimension of the Hilbert space, i.e., $1/D_{\cal H}$, in case $(\alpha)$ thermalization occurs later. Nonetheless, in case $(\beta)$, where the ETH condition is violated, although $w_s(0)$ approaches equilibration in the long time limit, it is not thermalized.

\begin{figure}
    \centering
    \includegraphics[width=0.3\textwidth]{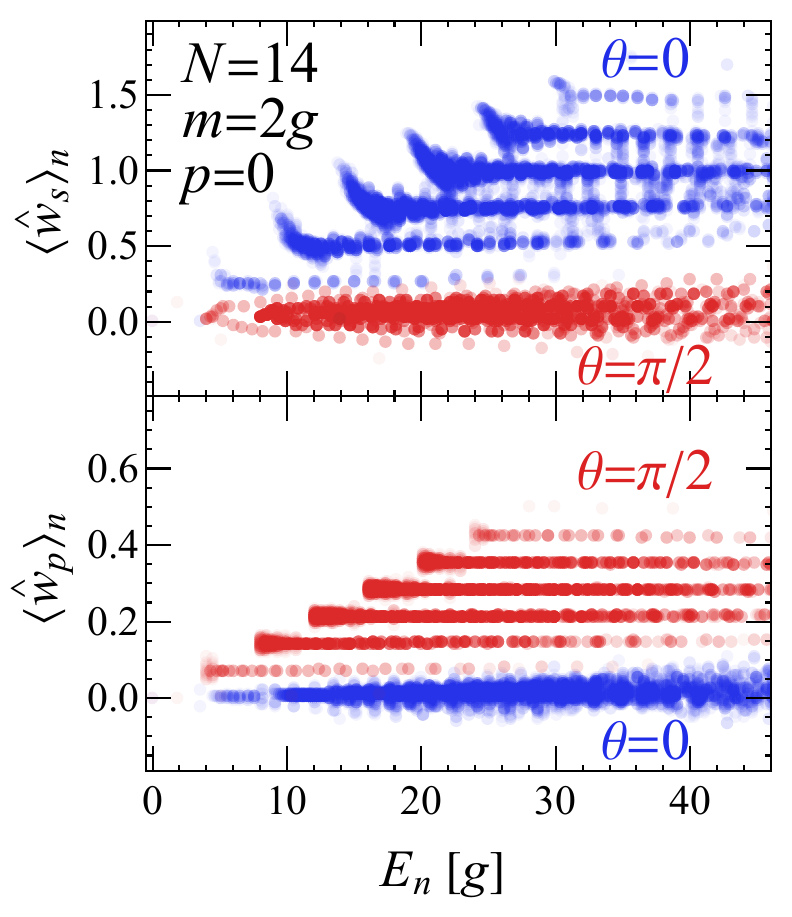}
    \caption{Comparison of ETH violation in the massive case ($m=2g$) with different $\theta$ angles.
    Upper and lower panels respectively present the scalar and pseudoscalar Wigner function at $p=0$ measured at different energy eigenstates. In each panel, blue (red) dots represent the case with $\theta=0$ ($\theta=\pi/2$). 
    \label{fig:eth_theta}}
\end{figure}

\vspace{3mm}
\textit{Violation of ETH, QMBS and $\theta$-vacuum.} 
The violation of ETH condition in an isolated non-integrable quantum system can be induced by the presence of quantum many-body localization (QMBL), or the existence of quantum many-body scar (QMBS) in the energy eigenstates. Particularly, although it has been noticed that the effect of disorder-free localization arises as a consequence gauge invariance in a Schwinger-like model~\cite{Smith_2017, Brenes:2017wzd, Papaefstathiou:2020rvv} \footnote{Localization in the Schwinger model with large mass is observed as well in~\cite{Ikeda:2023vfk} in the context of charge propagation.}, the gradual breaking down of the ETH in the scalar and the vector charge components of the Wigner function, as quark bare mass increases from strong to the weak coupling scenario, is largely associated with QMBS~\cite{Desaules:2022ibp}. 

The effect of QMBS is more transparent in the weak coupling limit, i.e. $m/g\to \infty$, where the influence of gauge field is substantially suppressed and the Hamiltonian effectively reduces to that of a massive fermion chain without interaction, $\hat H_{\rm fermion}$. The fermion chain contains a set of equally spaced eigenstates. One is allowed to define the operator, $\hat Q^\dagger_\alpha$,

\begin{equation}
    {\hat Q}^\dagger_\alpha \equiv \frac{1}{N_{\rm perm.}}\sum_{\substack{\rm perm. \\ |\mathcal{A}|+|\mathcal{B}| = \alpha}}
     \left(\prod_{j\in \mathcal{A}}
    \chi^\dagger_{j}\otimes 
    \prod_{k\in \mathcal{B}}
    \chi_{k}^{}\otimes 
    \prod_{i \in \overline{\mathcal{A} \cup \mathcal{B}}} I_i
    \right)\,,
\end{equation}
for $\alpha=1,2,\ldots$.
The summation of permutations includes all possible configurations of fermion (antifermion) creation operators $\chi^\dagger_j$ ($\chi_k^{}$) located on even (odd) lattice sites, $\mathcal{A}$ ($\mathcal{B}$) is the set of excited fermion (antifermion) sites with the number of elements being $|\mathcal{A}|$ ($|\mathcal{B}|$). Correspondingly, the complement of their union, $\overline{\mathcal{A} \cup \mathcal{B}}$, consists of the sites that are not occupied. $N_{\rm perm}$ is a normalization constant. Given the property that $[\hat H_{\rm fermion}, \hat Q^\dagger_\alpha] = \alpha \, m \, \hat Q_\alpha^\dagger$, with respect to the ground state that all of the fermion/antifermion sites are unoccupied $\ket{\Phi}=\ket{1010\ldots }$ and $\ket{Q_\alpha}=\hat Q_\alpha^\dagger\ket{\Phi}$, one has the tower of eigen-energy spectrum,
\begin{equation}
    \hat H_{\rm fermion} \ket{Q_\alpha} = (E_0+\alpha\, m)  \ket{Q_\alpha} \,.
\end{equation}
These eigenstates are the athermal QMBS states embedding along with the thermal eigenstate spectrum. Even if the fermion chain is coupled to the gauge field, the athermal subspace spanned by $\ket{Q_\alpha}$ persists with correction of order $g/m$ in the eigen-energy spacing, namely, $\hat H \ket{Q_\alpha}=(E_0+\alpha\, m) (1+O(g/m))\ket{Q_\alpha}$.   

Importantly, the athermal subspace from $\ket{Q_\alpha}$ is parity (P)-even, hence it does not affect the thermalization properties of P-odd operators, such as the pseudo-scalar components of the Wigner function. For instance, since $\bra{Q_\alpha}\hat w_p\ket{Q_\alpha'}=0$ as a consequence of the parity condition, the ETH condition in the eigenstate components of $(\hat w_p)_{mn}$ is not violated, in consistency with our observation. 
Moreover, the QMBS statement applies even if the parity property in the Hamiltonian is altered, e.g., by selecting a different $\theta$-vacuum. 
We thus choose $\theta=\pi/2$ in the Schwinger Hamiltonian~\eqref{eq:Hamiltonian} and \eqref{eq:Hamiltonian_lat} and the mass term becomes $m\bar \psi \gamma_5\psi$. Accordingly, the athermal subspace associated with the QMBS becomes parity-odd. Indeed, as observed in our numerical verification shown in Fig.~\ref{fig:eth_theta}, the thermalization properties in $\hat{w}_s$ and $\hat{w}_p$ swap.

\section{Quantum Thermalization in terms of entropy production}
\label{sec:entropy}

\begin{figure}[!hbtp]
    \centering
    \includegraphics[width=0.4\textwidth]{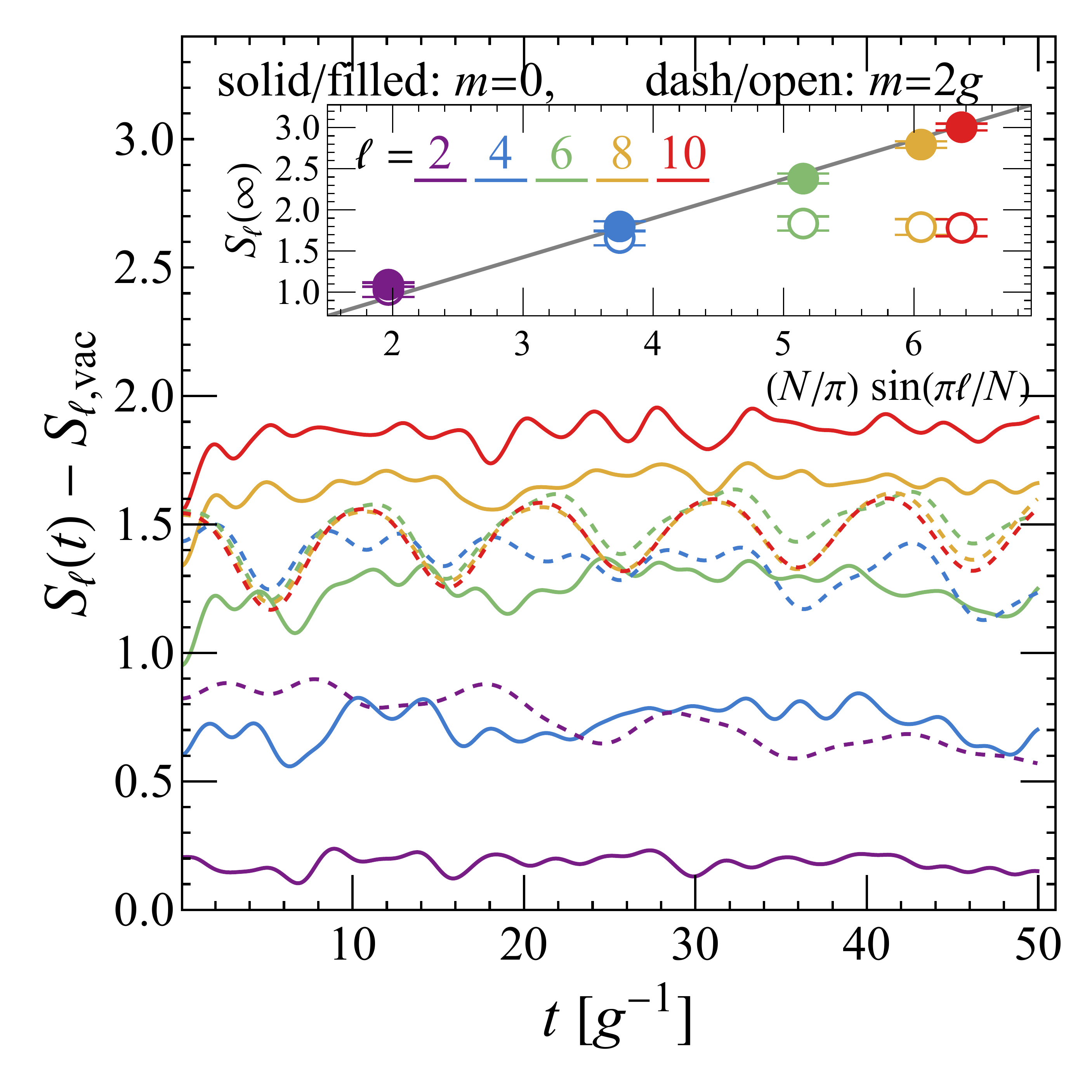}
    \caption{
    Time evolution of entanglement entropy with the corresponding vacuum value subtracted. In the inserted figure, their corresponding long-time limits are plotted with respect to the subsystem length with finite size correction.
    Solid curves and filled circles correspond to $m=0$, whereas dashed curves and open circles are for $m=2g$.
    Different colors represent subsystem size with $\ell=2$ (purple), $4$ (blue), $6$ (green), $8$ (orange), and $10$ (red), respectively.
    \label{fig:entropy}}
\end{figure}

It is instructive to check whether and how the quark system described by the Schwinger model approaches thermal equilibrium in terms of entropy. 
In this section, we compute the quantum entanglement entropy and classical Boltzmann entropy for the time-dependent quantum states used previously in computing the Wigner functions, i.e., in Fig.~\ref{fig:wigner} and~\ref{fig:wigner_diff}.

For a system with $N$ sites, one may split the system into two parts --- respectively denoted as $A$ and $\bar{A}$ and with length $\ell$ and $N-\ell$ --- and express a $2^{N}$-dimensional Hilbert space as the direct product of a $2^{\ell}$-dimensional subspace and a $2^{N-\ell}$-dimensional one.
For a time evolving pure state, $\ket{\Psi(t)}$, we compute the reduced density matrix $\rho^{A}(t) \equiv \mathrm{tr}_{\bar{A}}(\ket{\Psi(t)}\!\!\bra{\Psi(t)})$ and solve its eigenvalues for different $t$, denoted as $\lambda_i(t)$. We then compute the entropy of the reduced density matrix as $S(t) = -\sum_i\lambda_i\ln\lambda_i$, which is also known as the entanglement entropy between $A$ and $\bar{A}$.

\begin{figure}
    \centering
    \includegraphics[width=0.45\textwidth]{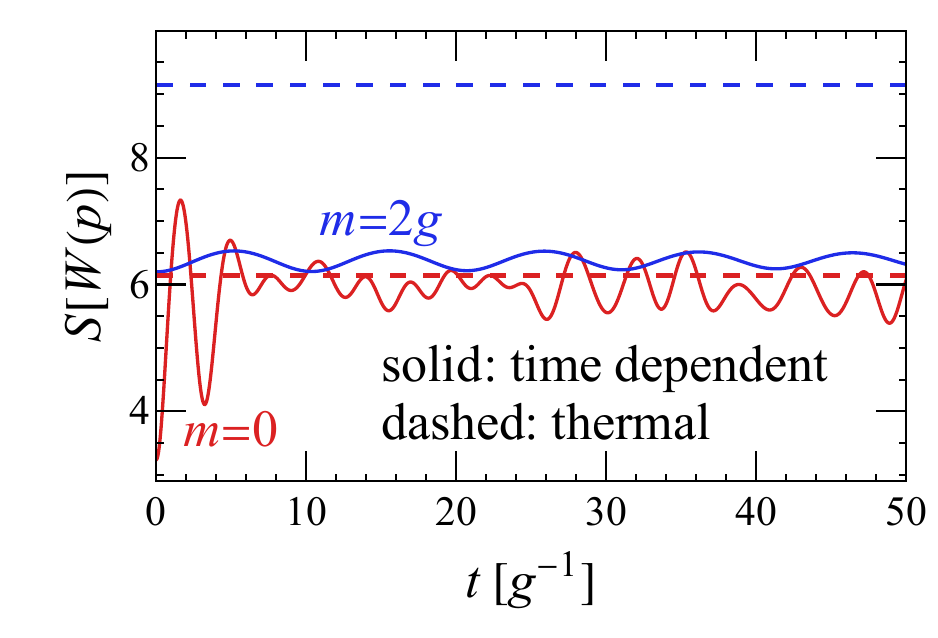}
    \caption{Boltzmann entropy as function of time. 
    \label{fig:Boltzmann_entropy}}
\end{figure}

We check the time evolution of the entropy for various subsystem sizes $\ell$, as shown in Fig.~\ref{fig:entropy}. For better comparison, the corresponding entropy measured for the ground state of the Hamiltonian, $S_{l,vac}$, has been subtracted for each curve. When $m=0$, $S_{\ell}$'s increase in time\footnote{Except for $\ell \leq 4$ which are dominated by quantum fluctuation.} and saturate after $t \sim 2g^{-1}$, then they fluctuate around the saturation value. 
Meanwhile, the long time limit of $S_{\ell}$ --- with mean and uncertainty respectively estimated by the average value and standard deviation in the time interval $t\in[10,50]g^{-1}$ --- is proportional to the subsystem length with finite size correction, $V(\ell) \equiv (N/\pi) \sin(\pi\ell/N)$, indicating that the entanglement entropy can be interpreted as the thermodynamic one. The volume can be intuitively understood by noting that $V(\ell) \approx \ell$ when $\ell\ll N$, and $V(\ell)=V(N-\ell)$ which shows the invariance of entropy when swapping the subsystem and its complement. One may refer to Ref.~\cite{Calabrese:2009qy} for more details.
In contrast, the $m=2g$ case does not exhibit a volume dependence or a saturation in time.

For comparison with the classical concept in the thermalization process, we also estimate the Boltzmann entropy. Following the standard Boltzmann prescription, we compute the quark and anti-quark distribution functions from the Winger functions, $f_{q,k}(t)=\frac{\langle \hat w_{s,k}+\hat w_{0,k}\rangle_t}{2}$ and $f_{\bar{q},k}(t)=\frac{\langle \hat w_{s,k}-\hat w_{0,k}\rangle_t}{2}$, respectively. Accordingly, the Boltzmann entropy is introduced as
\begin{align}
\begin{split}
    S[W(p)] \equiv\;&
    - \sum_{k=1}^{N} \Big( f_{q,k}\ln f_{q,k} + (1-f_{q,k})\ln (1-f_{q,k}) 
\\&
    + f_{\bar{q},k}\ln f_{\bar{q},k} + (1-f_{\bar{q},k})\ln (1-f_{\bar{q},k}) \Big)\,.
\end{split}
\end{align}
Note in Schwinger model, the gauge field is purely classical, there are no photon (gluon) excitations that contribute to the production of Boltzmann entropy.

In a system that reaches thermal equilibrium, the classical Boltzmann entropy should correspond to a distribution that approximates the canonical distribution, enabling the prediction of the equilibrium entropy of the quark system based on its initial energy.
In Fig.~\ref{fig:Boltzmann_entropy}, we present $S[W(p)]$ for both the strong and the weak coupling cases, with their corresponding thermal equilibrium expectations indicated as the dashed lines.
We observe the raising of $S[W(p)]$ in the strong coupling case for $t \lesssim 10/g$, reflecting the system's progression toward equilibrium, and then the entropy saturates, oscillating around the thermal equilibrium value. In contrast, in the weak coupling case, the Boltzmann entropy exhibits periodic oscillations and fails to converge to the thermal equilibrium value, indicating a lack of thermalization. 

The time dependence of both the entanglement entropy and the Boltzmann entropy, along with the subsystem size dependence observed in the former, consistently indicates thermalization in the strong coupling limit but not in the weak coupling case. This behavior aligns with the trends observed in the scalar component of the Wigner function. However, neither the entanglement entropy nor the Boltzmann entropy effectively captures the quantum thermalization of the pseudo-scalar component, highlighting the limitations of entropy in fully describing the quantum thermalization dynamics of all system observables. 

\section{Summary and Discussion}
\label{sec:summary}

In this work, we simulate the quantum thermalization process in the quark-gluon plasma. Using a non-perturbative approach based on the quantum computation algorithm, we solve the real-time evolution of a strongly coupled system formulated via the lattice Schwinger model and measure the time-dependent expectation values of the Wigner function operators. As quantum analogues of the phase-space distribution function, the thermalization of Wigner functions is studied in the context of time evolution and the verification of the eigenstate thermalization hypothesis.

In the strong-coupling limit, all components of the Wigner function successfully thermalize. In contrast, in the weak-coupling regime, only the pseudoscalar and axial-vector charge components of the Wigner function approach thermal equilibrium, while the scalar and vector charge components do not. This difference in thermalization behavior with respect to coupling strength is also reflected in the evolution of quantum entanglement entropy and classical Boltzmann entropy.

Additionally, the non-thermalization of the scalar and axial-vector components in the weak-coupling case is attributed to quantum many-body scar states, linked to the parity even mass term. A direct consequence of this is the unique influence of the non-trivial topological vacuum on quantum thermalization. Notably, when the system is promoted to a parity odd configuration with 
$\theta=\pi/2$, we have verified that the thermalization properties of the scalar and pseudo-scalar components are swapped.
This finding highlights the intricate interplay between coupling strength, intrinsic quantum states, and topological properties in governing the thermalization dynamics of quark-gluon plasma systems.

Although caution should be taken when extrapolating the current study to realistic quark-gluon plasmas—particularly in the context of QCD in 3+1 dimensions—there are several noteworthy implications that merit discussions.

First, our setup of lattice system with $N=20$ lattice sites provides at most a ten fermions and ten anti-fermions --- together with unlimited amount of photons (gluons) --- quark-gluon plasma. It provides insights of the real-time thermalization process of a small, isolated quantum many-body system governed by strong interaction.
Second, although the study with varying ratio $m/g$ reflects the physics of strong or weak couplings, it can as well be interpreted as the measure of quark bare mass relative to the system energy scale, as we have fixed the lattice size by the coupling constant, $ag=1/2$. Correspondingly, for instance, the non-thermalization of the scalar component of the Winger function in the cases of large ratio $m/g\sim ma$ indicates also the lack of thermalization of heavy quarks, such as charm quarks in a low-energy collision. 
Third, in the strong-coupling limit, the first excited state of the Hamiltonian corresponds to a meson with mass $M_\mathrm{meson} = g/\sqrt{\pi}$~\cite{Schwinger:1962tp}, and it corresponds to the lightest meson --- i.e., pion with mass $M_\pi = 135~\text{MeV}$ --- in the real world. Thermalization of the Wigner function and entropy occurs at $t \sim 2\,g^{-1} \approx 3\,M_\mathrm{meson}^{-1}$, which corresponds to $t\sim 3\,M_\pi^{-1} \approx 4~\mathrm{fm}/c$. This is in the same order of magnitude as the starting time ($\sim 1~\mathrm{fm}/c$) of hydrodynamic evolution determined phenomenologically by fitting the experimental data~\cite{JETSCAPE:2020shq, JETSCAPE:2020mzn, Nijs:2020roc, Nijs:2020ors}.

Last but not least, in the weak-coupling / massive scenario, we observe the distinct quantum thermalization behavior between parity even and parity odd operators, corresponding to the topological $\theta$-vaccum of the Schwinger model. Because in QCD the non-trivial topological vacuum arises analogously in the non-Abelian gauge field in terms of a $\theta$-term, together with a $\theta$ parameter that is expected extremely small (the strong CP), it is tempting to speculate that quantum thermalization should be easily achieved in the parity-odd observables, regardless of system energy scales, quark mass, etc. Indeed, such a statement is consistent with the remarkable success of the Statistical Hadronization Model (SHM)~\cite{Andronic:2009sv, Redlich:2009xx, Becattini:2009sc, Andronic:2017pug, He:2019tik}. As it applies to high-energy collisions from $e^+e^-$ to heavy ions, the final-state multiplicity distributions of pseudo-scalar meson states have been found compatible with the thermal equilibrium predictions. A detailed analysis of the role of the topological vacuum in QCD in quantum thermalization will be reported in our upcoming publication~\cite{CSYpapar:2025}.

\acknowledgements
The authors thank Xingyu Guo, Weiyao Ke, Ziwei Lin, and Yi Yin for helpful discussions.
This work is supported in part by Tsinghua University under grant Nos. 04200500123, 531205006, 100005024 (S.C. and S.S.) and in part by National Natural Science Foundation of China through Nos. 12375133 and 12147101 (L.Y.)

\begin{appendix}

\section{Schwinger Hamiltonian in a lattice with periodic boundary condition}\label{sec:app:schwinger_model}
In this Appendix, we provide a detailed introduction to how a lattice Schwinger Hamiltonian, with periodic boundary condition, is derived and represented by matrices. One may find similar discussions in many literature. However, in this work we represent the Hamiltonian with keeping the gauge field operator only at one site and employ the Gauss' law to fix others, which significantly reduces the volume of quantum states and allows one to perform simulation for larger lattices given the constraints of finite memory. We find it helpful to discuss the derivation in a systematic manner. 
In this work, we use $X$, $Y$, and $Z$ to represent the Pauli matrices, the metric convention with the temporal component being positive, $g^{\mu\nu} = g_{\mu\nu} = \mathrm{diag}(+1,-1)$, and the gamma matrices in $1+1$D is taken to be $\gamma^0 = Z$, $\gamma^1=iY$, and $\gamma^5 = \gamma^0 \gamma^1 = X$.

We start from the Schwinger Lagrangian density in the absence of $\theta$ term,
\begin{align}
\begin{split}
\mathcal{L} =\;& 
    \bar{\psi}\big(\gamma^{\mu}(i\partial_\mu - g\,A_\mu)-m\big)\psi
    -\frac{F^{\mu\nu} F_{\mu\nu}}{4}.
\end{split}
\end{align}
The grand-canonical momentum conjugate to the gauge ($A_\mu$) and spinor ($\psi$) fields are
\begin{align}
\mathcal{E}_0 \equiv\;&
    \frac{\delta \mathcal{L}}{\delta \dot{A}_0} = 0\,,\\
\mathcal{E} \equiv\;&
    \frac{\delta \mathcal{L}}{\delta \dot{A}_1} = \partial_t A_1 - \partial_z A_0\,,\\
\pi_\psi \equiv\;&
    \frac{\delta \mathcal{L}}{\delta (\partial_t \psi)} = i\,\psi^\dagger\,,
\end{align}
and the Hamiltonian density becomes
\begin{align}
\begin{split}
\mathcal{H} \equiv\;&
    \mathcal{E} \partial_0{A}_1 + \pi_\psi \partial_t \psi - \mathcal{L}\\
=\;&
    \bar{\psi}\big(\gamma^1 (-i \partial_z + g A_1) + m\big)\psi 
    + \frac{\mathcal{E}^2}{2}
\\&\,
    + \big(g\,\bar{\psi} \gamma^0 \psi
     - \partial_z\mathcal{E} \big)A_0.
\end{split}\label{eq:ham_0}
\end{align}
In what follows, we take the Schr\"odinger picture that operators does not depend on time, and the commutation and anticommutation relations between fields and their corresponding conjugate momenta are $[A_1(z), \mathcal{E}(z')] = i\, \delta(z-z)$
and $\{\psi_a(z), \pi_\psi^b(z')\} = i\,\delta_a^b\,\delta(z-z')$.

Under a gauge transformation,
\begin{align}
    A_\mu \to A_\mu - g^{-1}\partial_\mu \varphi,
\qquad
    \psi \to e^{i\varphi} \psi,
\label{eq:gauge_transformation}
\end{align}
we find the Lagrangian density is invariant, $\mathcal{L} \to \mathcal{L}$, so should be Hamiltonian. Nevertheless, one can find that under~\eqref{eq:gauge_transformation}, $\mathcal{H} \to \mathcal{H} - (\partial_z\mathcal{E} - g\, \bar{\psi} \gamma^0 \psi)\partial_t{\varphi}$, which implies that the Gauss' law
\begin{align}
    \partial_z\mathcal{E} = g\, \bar{\psi} \gamma^0 \psi
\end{align}
shall be explicitly implemented to maintain Gauge invariance.
The second line in the Hamiltonian~\eqref{eq:ham_0} vanishes, and the $\mathcal{E}$ field at different positions shall no longer be regarded as independent with each other. They are given by 
\begin{align}
    \mathcal{E}(z) = \mathcal{E}_\mathrm{bnd} + g \int_{z_0}^z \bar{\psi}(z') \gamma^0 \psi(z') \mathrm{d}z'\,,
\end{align}
where $z_0$ is the lower boundary in $z$, and $\mathcal{E}_\mathrm{bnd}$ is the electric field operator at $z_0$.
We consider the Schwinger model defined in a finite interval $z\in[0,L]$, with the periodic boundary condition that $\psi(L) = \psi(0)$ and $\mathcal{E}(L) = \mathcal{E}(0) = \mathcal{E}_\mathrm{bnd}$. The latter requires that the total charge of the system must be vanishing, $0=\int_{0}^L \bar{\psi}(z) \gamma^0 \psi(z) \mathrm{d}z$. 

We further perform a gauge transformation with phase being 
\begin{align}
    \varphi(z) = g\int_{0}^z A_1(z') \mathrm{d}z',
    \label{eq:gauge_fixing}
\end{align}
so that $\bar{\psi}\gamma^1 (-i \partial_z + g A_1)\psi \to -i \bar{\psi}\gamma^1 \partial_z\psi$, and the boundary condition becomes $\psi(0) = e^{-i\,g\int_{0}^{L} A_1(z) \mathrm{d}z} \psi(L)$.
The Hamiltonian reads
\begin{align}
    H = \int_{0}^{L} 
    \Big( \bar{\psi}\big(-i\gamma^1 \partial_z + m\big)\psi
    + \frac{1}{2}\mathcal{E}^2 \Big)\mathrm{d}z\,.
\end{align}

We discretize the $z$ direction into lattice with spacing $a$, fields are defined on the grids $z_n = n\,a$, for $n = 1, \cdots, N$. For fermions, we use the staggered fermion prescription which was introduced by Kogut and Susskind~\cite{Kogut:1974ag, Susskind:1976jm}, 
\begin{align}
    \chi_{2n}^{} = a^{\frac{1}{2}}\psi_{\uparrow}(z_{2n})\,,\qquad
    \chi_{2n+1}^{} = a^{\frac{1}{2}}\psi_{\downarrow}(z_{2n+1})\,.
\end{align}
They follow that anticommutation relations
\begin{align}
\begin{split}
\{ \chi^\dagger_n, \chi^{}_m \} = \delta_{nm}\,,\quad
\{ \chi^\dagger_n, \chi^\dagger_m \} = \{ \chi^{}_n, \chi^{}_m \} = 0\,.
\end{split}
\label{eq:fermion_anticommutation}
\end{align}
There is one remaining independent gauge field operator --- the one at the boundary,
\begin{align}
\varepsilon = g^{-1} \mathcal{E}_\mathrm{bnd}, \qquad
U = e^{-i\,g\int_{0}^{L} A_1(z) \mathrm{d}z},
\end{align}
and they satisfy $[\varepsilon, U] = -U$.
With these, the lattice Schwinger Hamiltonian reads,
\begin{align}
\begin{split}
H_\mathrm{lat}
=\;&
    -\frac{i}{2a} \sum_{n=1}^{N-1} 
    \Big(  \chi^\dagger_{n} \chi^{}_{n+1} - \chi^\dagger_{n+1}  \chi^{}_{n}\Big)
\\&
    -\frac{i}{2a} 
    \Big(\chi^\dagger_{N} U^\dagger \chi^{}_{1} - \chi^\dagger_{1} U \chi^{}_{N}\Big)
\\&
    +\sum_{n=1}^{N}\Bigg( (-1)^n m\,\chi_n^\dagger \chi^{}_n
    + \frac{a\, g^2}{2}\varepsilon_n^2\Bigg)\,,
\end{split}
\label{eq:Hamiltonian_lat_0}
\end{align}
with the electric field given by
\begin{align}
    \varepsilon_n = \varepsilon + :q_n: = \varepsilon + \sum_{m=1}^n (\chi_m^\dagger \chi_m^{} - \frac{1-(-1)^m}{2})\,,
\end{align}
and $::$ denotes a normal product. In the presence of finite $\theta$ term, \eqref{eq:Hamiltonian_lat_0} becomes Eq.~\eqref{eq:Hamiltonian_lat} in the main text.

In the lattice Schwinger Hamiltonian, $\chi_n^{}$, $\chi_n^\dagger$, $\varepsilon$ and $U$ are operators defined by the desired commutation and/or anti-commutation relations. The spinor operators can be realized by the Jordan--Wigner representation~\cite{Jordan:1928wi},
\begin{align}
\begin{split}
    \chi^{}_n =\;& \frac{X_n-iY_n}{2}\prod_{m=1}^{n-1}(-i Z_m),\\
    \chi^\dag_n =\;& \frac{X_n+iY_n}{2}\prod_{m=1}^{n-1}(i Z_m),
\end{split}
\end{align}
where we have used the notation $X_n \equiv \big(\prod_{j=1}^{n-1}\otimes I\big) \otimes X \otimes \big(\prod_{j=n+1}^{N}\otimes I\big)$ and likewise for $Y_n$ and $Z_n$.
The electric-filed operator and the link operator reads
\begin{align}
\varepsilon &=
    \sum_{\epsilon=-\Lambda}^{\Lambda}\epsilon\ket{\epsilon}\bra{\epsilon},\\
U&=
    \sum_{\epsilon=-\Lambda}^{\Lambda-1}\ket{\epsilon}\bra{\epsilon+1},
\quad
    U \ket{-\Lambda} = \ket{\Lambda},
    \label{eq.mat_U}
\end{align}
where $\ket{\epsilon}$'s are the eigenbasis of electric field operator $\varepsilon$, and $\Lambda$ is a cutoff that truncates the infinite dimensional basis~\cite{Shaw:2020udc}.
Implementing the matrices representation of the fields, the lattice Hamiltonian~\eqref{eq:Hamiltonian_lat} becomes
\begin{align}
\begin{split}
H_\mathrm{gate}
=\;&
     \frac{m}{2} I_G\otimes \sum_{n=1}^{N} (-1)^n\, Z_n
\\&+
    \frac{1}{4a}I_G\otimes \sum_{n=1}^{N-1} (X_n X_{n+1}+Y_n Y_{n+1})
\\&+
    \frac{1}{8a}\Big(
    (U + U^\dagger)\otimes(X_N X_{1} + Y_{N}Y_{1})
\\&\quad\;+
    i(U - U^\dagger)\otimes(X_N Y_{1} - Y_{N}X_{1})\Big)
\\&
     + \frac{a\, g^2}{2} \sum_{n=1}^{N} \Big(\varepsilon \otimes I_F + I_G\otimes \sum_{j=1}^{n} \frac{Z_j+(-1)^j}{2}\Big)^2\,,
\end{split}
\end{align}
where $I_G$ and $I_F$ are respectively the identity matrices in the gauge and fermion fields' Hilbert space.

Now we move on to construct the equal-time Wigner operators in the Lattice theory. The Wigner operator's are $2\times2$-dimensional matrices, with the $(a,b)$-th element given by [note that the gauge links become identity after taking the gauge fixing~\eqref{eq:gauge_fixing}]
\begin{align}
    \hat{W}_{ab}(z,p) \equiv \int_{-L}^{L} \bar\psi_a(z_+)\, \psi_b(z_-)\, e^{i\, p\, y}\,dy\,,
\end{align}
where $z_\pm \equiv z\pm\frac{y}{2}$.
By tracing out the Dirac indices, one may decompose them into scalar, vector, axial vector, and pseudoscalar sectors, 
\begin{align}
\begin{split}
\hat{w}_s \equiv\;& 
    \frac{\mathrm{tr}(\hat{W})}{2} 
    = \int
    \bar\psi(z_+)\, \psi(z_-)\, e^{i\, p\, y}\,\frac{dy}{2}\,,\\
\hat{w}_p \equiv\;& 
    \frac{\mathrm{tr}(i\gamma^5 \hat{W})}{2} 
    = i\int
    \bar\psi(z_+)\,\gamma^5 \psi(z_-)\, e^{i\, p\, y}\,\frac{dy}{2}\,,\\
\hat{w}_0 \equiv\;& 
    \frac{\mathrm{tr}(\gamma^0 \hat{W})}{2} 
    = \int
    \psi^\dagger(z_+)\, \psi(z_-)\, e^{i\, p\, y}\,\frac{dy}{2}\,,\\
\hat{w}_1 \equiv\;& 
    -\frac{\mathrm{tr}(\gamma^1 \hat{W})}{2} 
    = \int
    \bar\psi(z_+)\,\gamma^1 \psi(z_-)\, e^{i\, p\, y}\,\frac{dy}{2}\,,\\
\end{split}
\end{align}
so that 
\begin{align}
    \hat{W} = \hat{w}_s - i\,\hat{w}_p \gamma^5 + \hat{w}_0 \gamma^0 + \hat{w}_1 \gamma^1\,.
\end{align}
On a lattice, we replace the integration by summation over grids. That is, $\int_{-L}^{L} dy f(y) \to 2a\sum_{m=-\frac{N}{2}}^{\frac{N}{2}-1} f(2ma)$ for $\hat{w}_s$ and $\hat{w}_0$, 
\begin{align}
\begin{split}
    &\hat{w}_s(z_n,p) 
\\=\;&
    a\sum_{m=-\frac{N}{2}}^{\frac{N}{2}-1} e^{2i\,m\,a\,p}
    \times\Big(\psi_{\uparrow}^{\dagger}(z_{n+m}) \psi_{\uparrow}^{}(z_{n-m})
\\&\qquad\quad
    -\psi_{\downarrow}^{\dagger}(z_{n+m}) \psi_{\downarrow}^{}(z_{n-m})\Big)
\\=\;&
    \sum_{m=-\frac{N}{2}}^{\frac{N}{2}-1} e^{2i\,m\,a\,p}
    (-1)^{n+m} \chi^{\dagger}_{n+m} \chi^{}_{n-m}\,,
\end{split}
\end{align}
and likewise,
\begin{align}
    \hat{w}_0(z_n,p) = \sum_{m=-\frac{N}{2}}^{\frac{N}{2}-1} e^{2i\,m\,a\,p} \chi^{\dagger}_{n+m} \chi^{}_{n-m}\,.
\end{align}
While for $\hat{w}_1$ and $\hat{w}_p$, the replacement should read $\int_{-L+a}^{L+a} dy f(y) \to 2a\sum_{m=-\frac{N}{2}}^{\frac{N}{2}-1} f((2m+1)a)$. We find
\begin{align}
    \hat{w}_1(z_{n-\frac{1}{2}},p) =\;&
    \sum_{m=-\frac{N}{2}}^{\frac{N}{2}-1} e^{i\,(2m+1)\,a\,p} \chi^{\dagger}_{n+m} \chi^{}_{n-m-1}\,,
\end{align}
and
\begin{align}
\begin{split}
    &\hat{w}_p(z_{n-\frac{1}{2}},p) 
\\=\;&
    i\sum_{m=-\frac{N}{2}}^{\frac{N}{2}-1} e^{i\,(2m+1)\,a\,p} (-1)^{n+m}\chi^{\dagger}_{n+m} \chi^{}_{n-m-1}\,.
\end{split}
\end{align}
We note that the gauge fixing~\eqref{eq:gauge_fixing} in the current paper is equivalent to the discrete gauge transformation discussed in e.g., Ref.~\cite{Ikeda:2020agk}. In general gauge, one shall keep the gauge field operators, $U_n\equiv e^{-i\,a\,g\,A_1(z=n\,a)}$, and terms expressed as $\chi^{\dagger}_{n} \chi^{}_{m}$ shall return to $\chi^{\dagger}_{n} U_{n-1} \cdots U_{m+1} U_{m} \chi^{}_{m}$ when $n>m$, which resembles the expression of Wigner function with gauge link~\eqref{eq:wignerf}. 

Noting the periodic boundary condition of the spinor fields, the operators in the summation shall be invariant under the translation $m\to m+N$, which respectively means $e^{2i\,N\,a\,p} = 1$. Therefore, the Wigner operators shall be computed at $p = p_k \equiv \frac{\pi}{N a}k$. 
It can be shown that operators $\hat{w}_i(z,p_k)$ always vanish when $k$ is odd. We therefore focus on even sites in the momentum grids. It is also obvious to find that $\hat{w}_i(z,p) = \hat{w}_i(z,p+2N\frac{\pi}{N a})$ for all components. Meanwhile, one may further find $\hat{w}_i(z,p) = \hat{w}_i(z,p+N\frac{\pi}{N a})$ for $i\in\{s,0\}$ and $\hat{w}_i(z,p) = -\hat{w}_i(z,p+N\frac{\pi}{N a})$ for $i\in\{p,1\}$. 

\vspace{4mm}\textit{Properties of the Wigner functions.} --- 
The properties of the Wigner functions can be more clearly seen from expressing the field operators in the coordinate space by those in the momemtum space,
\begin{align}
\psi(z) =\;&
    \int \frac{dp}{2\pi\sqrt{2p^0}}\Big(
    u(p) \hat{f}_p e^{i p z} +  v(p) \hat{\bar{f}}^\dagger_p e^{-i p z}
    \Big)\,,\\
\bar{\psi}(z) =\;&
    \int \frac{dp}{2\pi\sqrt{2p^0}}\Big(
    \bar{u}(p) \hat{f}_p^\dagger e^{-i p z} +  \bar{v}(p) \hat{\bar{f}}_p e^{i p z}
    \Big)\,,
\end{align}
with $p^0 = \sqrt{p^2+m^2}$, $\hat{\bar{f}}^\dagger_p$ ($\hat{\bar{f}}_p$) is the creation (annihilation) operator for an antifermion with momentum $p$ and likewise $\hat{f}^\dagger_p$ and $\hat{f}_p$ are for fermions,  and
\begin{align}
u(p) = 
\left(\begin{array}{c}
    \sqrt{p^0+m} \\
    p/\sqrt{p^0+m}
\end{array}\right)\,,\quad
v(p) = 
\left(\begin{array}{c}
    p/\sqrt{p^0+m} \\
    \sqrt{p^0+m}
\end{array}\right)\,.
\end{align}
The spatial averaged Wigner operators are expressed as 
\begin{align}
\begin{split}
\hat{w}_0(p) =\;&
    \frac{2}{L}\,
    \Big(\hat{f}_{p}^\dagger \hat{f}_{p}
    -\hat{\bar{f}}^\dagger_{-p} \hat{\bar{f}}_{-p}\Big)\,, 
\\
\hat{w}_p(p) =\;&
    \frac{i}{2 L}
    \Big(\hat{f}_{p}^\dagger \hat{\bar{f}}^\dagger_{-p}
    + \hat{f}_{p} \hat{\bar{f}}_{-p} \Big)\,,
\\
\hat{w}_s(p) =\;&
    \frac{m}{2\sqrt{m^2+p^2}L}
    \Big(\hat{f}_{p}^\dagger \hat{f}_{p}
    +\hat{\bar{f}}^\dagger_{-p} \hat{\bar{f}}_{-p}\Big)
\\&
    - \frac{p}{2\sqrt{m^2+p^2}L}
    \Big(\hat{f}_{p}^\dagger \hat{\bar{f}}^\dagger_{-p}
    - \hat{f}_{p} \hat{\bar{f}}_{-p} \Big),
\\
\hat{w}_1(p) =\;&
    \frac{p}{2\sqrt{m^2+p^2}L}
    \Big(\hat{f}_{p}^\dagger \hat{f}_{p}
    + \hat{\bar{f}}^\dagger_{-p} \hat{\bar{f}}_{-p}\Big)
\\&
    + \frac{m}{2\sqrt{m^2+p^2}L}
    \Big(\hat{f}_{p}^\dagger \hat{\bar{f}}^\dagger_{-p}
    - \hat{f}_{p} \hat{\bar{f}}_{-p} \Big)\,.
\end{split}\label{eq:wigner_momentum_operator}
\end{align}

Under parity ($\mathrm{P}$) and charge-conjugation ($\mathrm{C}$) transformations, the annihilation operators transform as
\begin{align}\begin{split}
    \mathrm{P} f_{p} \mathrm{P} = f_{-p}\,,&\quad
    \mathrm{P} \bar{f}_{p} \mathrm{P} = -\bar{f}_{-p}\,,\\
    \mathrm{C} f_{p} \mathrm{C} = \bar{f}_{p}\,,&\quad
    \mathrm{C} \bar{f}_{p} \mathrm{C} = f_{p}\,,
\end{split}\end{align}
and likewise for their corresponding creation operators. Therefore, the Wigner functions follow
\begin{align}\begin{split}
    \mathrm{P} \hat{w}_s(p) \mathrm{P} = \hat{w}_s(-p)
&\,,\quad
    \mathrm{C} \hat{w}_s(p) \mathrm{C} =\hat{w}_s(-p)
\,,\\
    \mathrm{P} \hat{w}_p(p) \mathrm{P} = -\hat{w}_p(-p)
&\,,\quad
    \mathrm{C} \hat{w}_p(p) \mathrm{C} = -\hat{w}_p(-p)
\,,\\
    \mathrm{P} \hat{w}_0(p) \mathrm{P} =\hat{w}_0(-p)
&\,,\quad
    \mathrm{C} \hat{w}_0(p) \mathrm{C} = -\hat{w}_0(-p)
\,,\\
    \mathrm{P} \hat{w}_1(p) \mathrm{P} = -\hat{w}_1(-p)
&\,,\quad
    \mathrm{C} \hat{w}_1(p) \mathrm{C} = -\hat{w}_1(-p)
\,.
\end{split}
\end{align}

\section{Thermal Expectation and Time Evolution of Observables}
\label{sec:app:state}
In this work, we computed the thermal expectation and the time evolution of Wigner function using a numerically efficient method. We first find the $N_\mathrm{trunc}$ lowest eigenvalues of the Hamiltonian matrix and the corresponding eigenstates,
\begin{align}
    \hat{H} \ket{E_n} = E_n \ket{E_n}\,,\qquad n=0,1,\cdots,N_\mathrm{trunc}-1.
\end{align}
Then, for a quantum state with initial condition, 
\begin{align}
    \ket{\Psi(t=0)} = \sum_{n=0}^{N_\mathrm{trunc}-1} c_n \ket{E_n},
\end{align}
its time evolution is given by 
\begin{align}
    \ket{\Psi(t)} = \sum_{n=0}^{N_\mathrm{trunc}-1} c_n\, e^{-i\,E_n t} \ket{E_n},
\end{align}
and the expectation value of an arbitrary operator ($O$) is known as
\begin{align}
    \langle \hat{O} \rangle_t \equiv 
    \bra{\Psi(t)} \hat{O} \ket{\Psi(t)} 
    =  \sum_{n,n'=0}^{N_\mathrm{trunc}-1}
    c_n^{} c_{n'}^* e^{i(E_{n'}-E_n)t} O_{n',n}
    \,,
\end{align}
where $O_{n',n} \equiv \bra{E_{n'}} \hat{O} \ket{E_n}$ is the matrix element.
For a realistic measurement with finite time resolution, what is measured is a time-averaged effect. We therefore define the long-time average 
\begin{align}
\begin{split}
    \overline{\langle O \rangle_t}
\equiv\;&
    \lim_{t\to\infty}\frac{1}{t}\int_0^t \langle O \rangle_{t'} \mathrm{d}{t'}\\
=\;&
    \sum_{n,n'=0}^{N_\mathrm{trunc}-1}
    c_n^{} c_{n'}^* \delta_{E_{n'},E_n}
    O_{n',n} 
    \,,
\end{split}
\end{align}
where $\delta_{E_{n'},E_n}$ is a Kronecker-$\delta$ symbol ensuring the summation only apply to states with $E_{n} = E_{n'}$, i.e., within the same state or between degenerated states.

The thermal equilibrium expectation can be calculated in two scenarios.
The microcanonical ensemble (MCE) average within the energy shell $E_n \in [E-\frac{\Delta E}{2}, E+\frac{\Delta E}{2}]$ is given by 
\begin{align}
    \langle \hat{O} \rangle_\mathrm{MCE}(E,\Delta E) = \frac{\sum_{n:\,|E_n -E| \leq \frac{\Delta E}{2}} O_{n,n}
    } {\sum_{n:\,|E_n -E| \leq \frac{\Delta E}{2}} 1}\,,
\end{align}
with the corresponding thermal fluctuation given by 
\begin{align}
    \delta_O^\mathrm{MCE} = \bigg(\frac{\sum_{n:\,|E_n -E| \leq \frac{\Delta E}{2}} O_{n,n}^2}
    {\sum_{n:\,|E_n -E| \leq \frac{\Delta E}{2}} 1} - \langle \hat{O} \rangle_\mathrm{MCE}^2\bigg)^{\frac{1}{2}}\,.
\end{align}
The canonical ensemble (CE) average and variance with temperature $T$ is given by
\begin{align}
    \langle \hat{O} \rangle_\mathrm{CE}(T) =\;& \frac{\sum_{n=0}^{N_\mathrm{trunc}-1} O_{n,n}
    e^{-E_n/T}}{\sum_{n=0}^{N_\mathrm{trunc}-1} e^{-E_n/T}}\,,\\
    \delta_O^\mathrm{CE}(T) =\;& \bigg(\frac{\sum_{n=0}^{N_\mathrm{trunc}-1} O_{n,n}^2
    e^{-E_n/T}}{\sum_{n=0}^{N_\mathrm{trunc}-1} e^{-E_n/T}} - \langle \hat{O} \rangle_\mathrm{CE}^2\bigg)^{\frac{1}{2}}\,,
\end{align}
where the temperature for a given quantum state shall be given by fixing the energy,
\begin{align}
\langle \hat{H} \rangle_\mathrm{CE}(T)
    = \bra{\Psi} \hat{H} \ket{\Psi}\,.
\end{align}

Therefore, for the observables of interest, we compute the matrix elements $O_{n',n}$, and the real-time evolution and thermal expectation values of $\hat{O}$ can be obtained in a straightforward and efficient manner.

\begin{figure}[!h]
    \centering
    \includegraphics[width=0.4\textwidth]{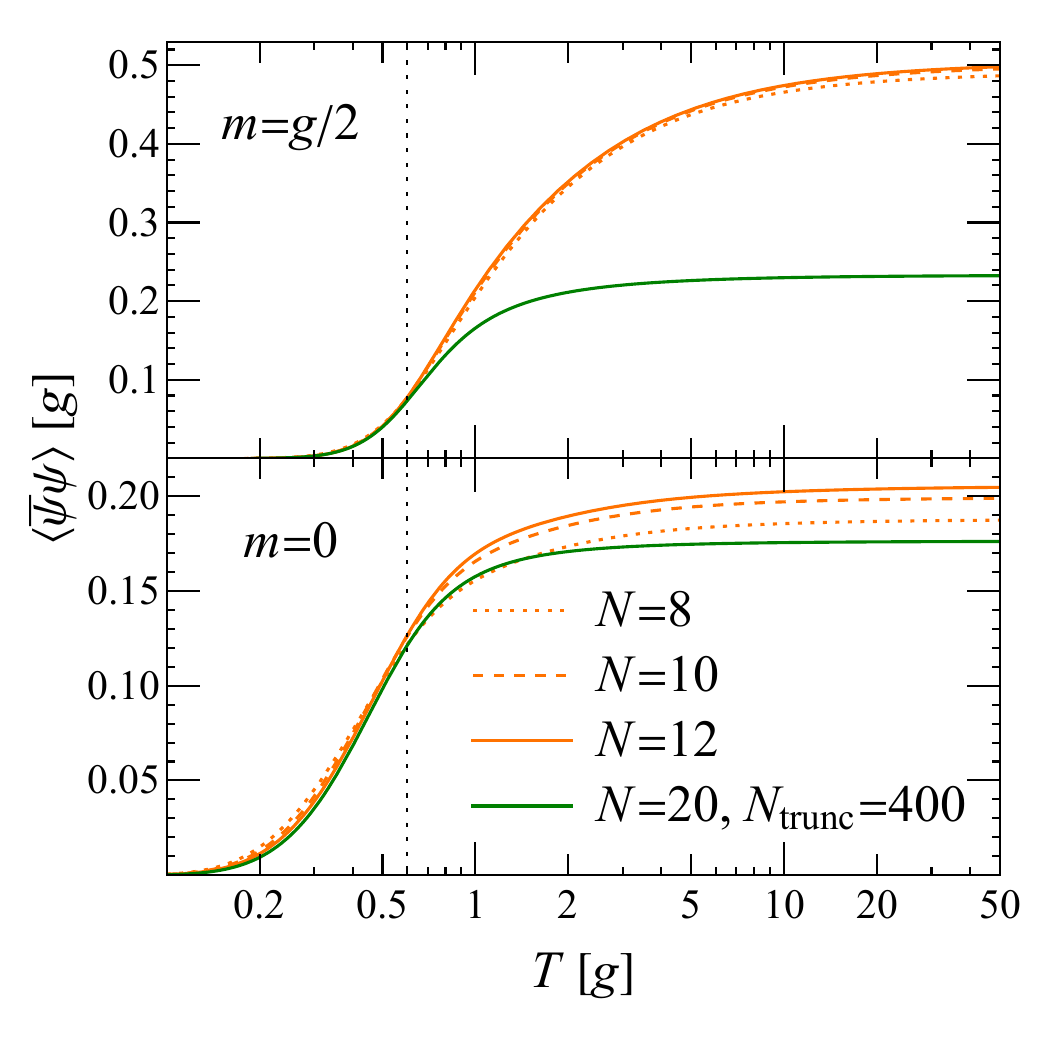}
    \caption{Temperature dependence of scalar condensate $\langle \bar\psi\psi\rangle$ from calculation in the truncated Hilbert space with $N_{\mathrm{trunc}}=400$ states in $N=20$(green), compared to those from calculation in the full Hilbert space in small systems with $N=8$(orange dotted), $10$(orange dashed), and $12$(orange solid). 
    A vertical line located at $T = 0.6g$ is added to indicate the upper bound of temperature that results with the truncated Hilbert space are consistent with full calculation with full Hilbert space.
    Upper and lower panel correspond to fermion mass $m=g/2$ and $m=0$, respectively. The gauge field truncation is taken as $\Lambda=2$ in all curves. 
    \label{fig:truncation}}
\end{figure}
For the parameter set up in the main text, $N=20$, $\Lambda=2$, we considered $N_\mathrm{trunc} = 400$ states. This allows us to study the thermal observables for systems with energy below $E_{N_\mathrm{trunc}}$ in MCE and for those with low temperature in CE, where the partition function is dorminated by low-energy excitations. In what follows we discuss the upper bound of CE temperature that one may trust. While it is computationally expensive to find all the energy eigenstates in a system with $N=20$, we keep $\Lambda=2$ and vary the lattice size to be $N=8$, $10$, and $12$ and diagonaize the Hamiltonian matrix completely. In the neutral sector (i.e., total charge $Q=0$), the dimensions of complete Hilbert space are $350$, $1260$, and $4620$, respectively. Taking into account all the quantum states, we compute the thermal expectation of scalar condensate in small system from low ($T=0.1g$) to high temperature $T=50g$ (see orange curves in Fig.~\ref{fig:truncation}), and then we compared them with the corresponding results in the truncated Hilbert space of a large system (green curve).
Calculations with full Hilbert space in small systems exhibit weak size dependence, which allows one to extrapolate the results to a large system. By comparing the green and orange curves, we observe that they are consistent for temperature $T \lesssim 0.6\, g$. Above such a temperature, CE results are quantitatively inconsistent, although they exhibit similar qualitative behaviors. Therefore, we conclude that within the truncated Hilbert space with $N_{\mathrm{trunc}} = 400$ states, CE results are reliable up to temperature $T\sim 0.6\, g$.

\section{More results on ETH}\label{sec:app:eth}
In this Appendix, we provide more systematic results on the ETH test.

\begin{figure}[!hbpt]
    \centering
    \includegraphics[width=0.23\textwidth]{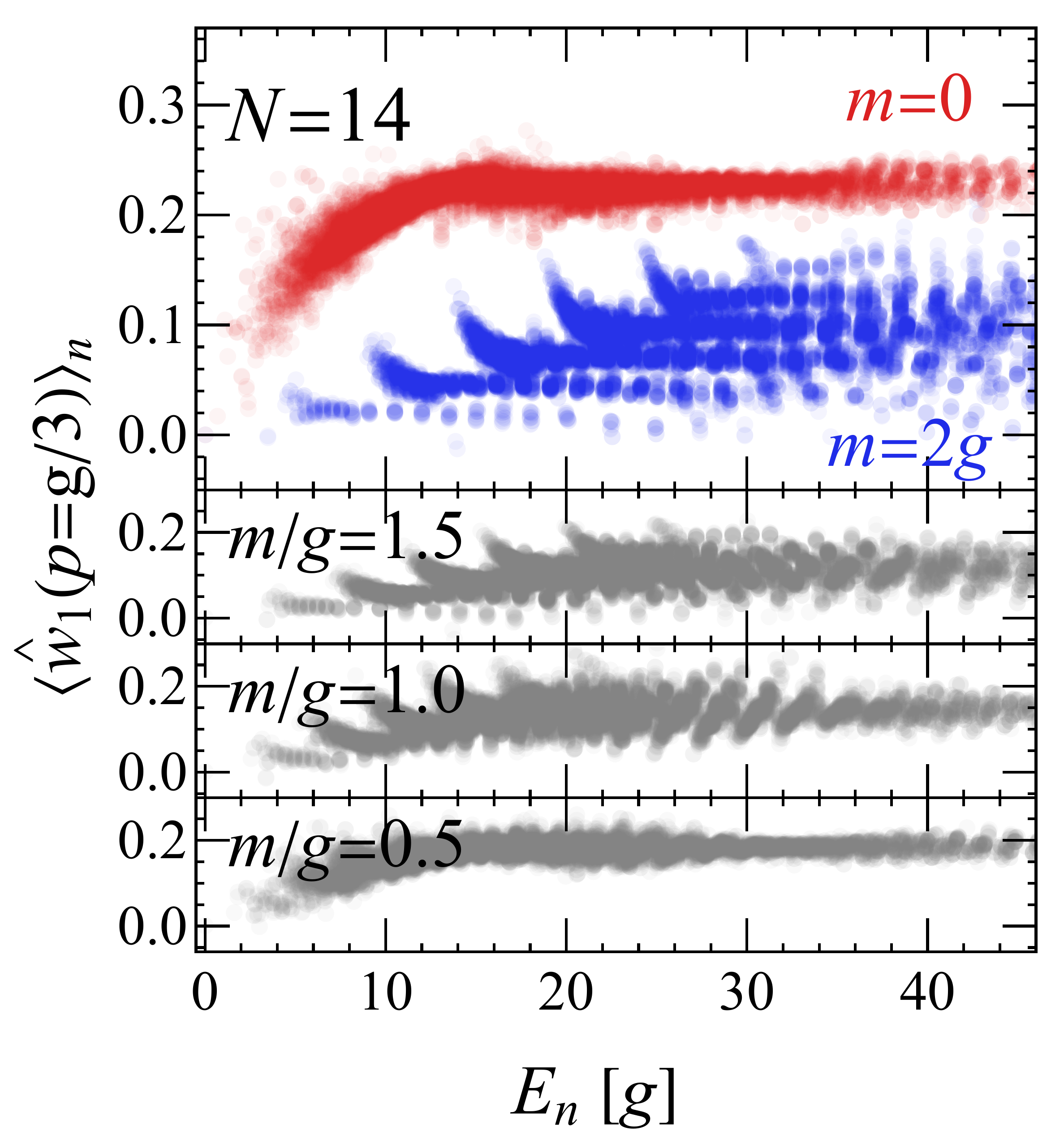}
    \includegraphics[width=0.23\textwidth]{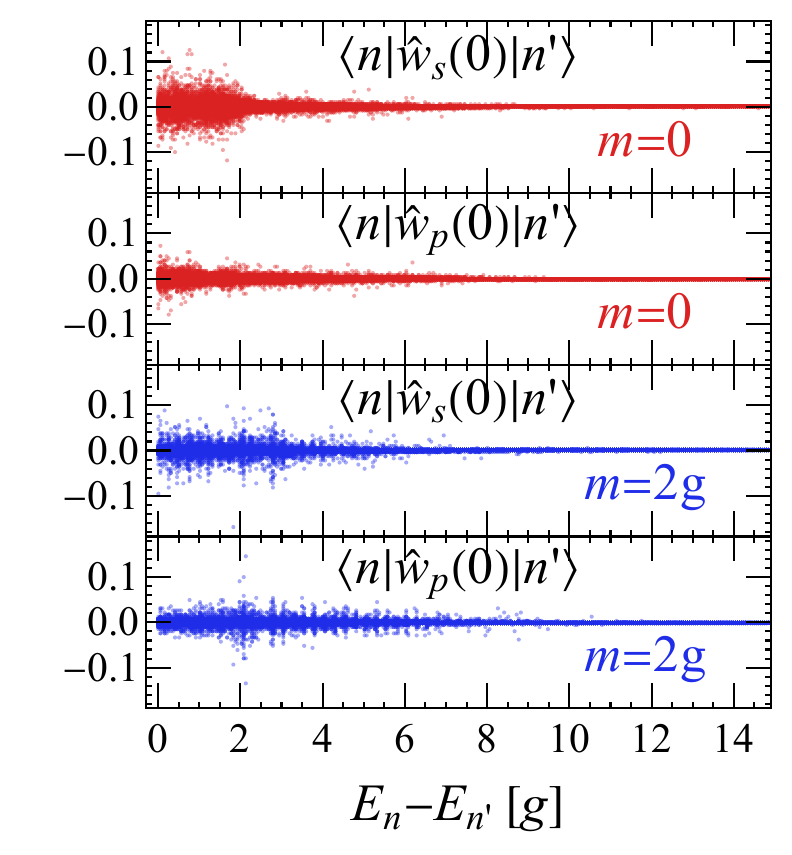}
    \caption{
    (Left) Same as Fig.~\protect{\ref{fig:eth}} in the main text but for axial charge component of  Wigner function. Vector components are not shown since $\langle \hat{w}_v \rangle_n$'s are always vanishing for all $n$, $p$, and $m$.
    (Right) Off-diagonal elements of the Wigner function matrix in the energy eigenstate basis. Horizontal axis represents the energy difference between two states. We only plot the positive energy differences as the matrix is Hermitian.
    We show the scalar and pseudoscalar components at $p=0$ as the representive, but other components at different momentum points share the same quantitative properties. The left panel corresponds to systems with $N=14$ sites, whereas the right panel is for $N=12$ owing to limitation of computation resources. We have used the notation $\langle \hat{O} \rangle_n \equiv \bra{E_n} \hat{O} \ket{E_n}$.
    \label{fig:eth2}}
\end{figure}
\textit{Other components in the Wigner function and off diagonal elements.} --- 
To examine to what extent the ETH~\eqref{eq:eth} is satisfied, we also check other components of the Wigner function (Fig.~\ref{fig:eth2}-left) and the off-diagonal elements in the Wigner function matrix (Fig.~\ref{fig:eth2}-right). We only present the results at $p=0$, but other momentum points share the same quantitative properties. As expected, such off-diagonal elements randomly distribute around zero, and the variance decreases with the energy difference between the two energy eigenstates. Keeping in mind that the time dependent Wigner functions are given by $\sum_{n,n'} e^{i(E_n-E_n')t}\bra{E_n} \hat{W} \ket{E_{n'}}$, the energy-difference dependence of the variance suppresses the high-frequency oscillation, whereas the zero-mean high-statistics samples in the small energy-difference sector ensures the cancellation between different low-frequency modes. Finally, the maximum width when $E_n-E_{n'}$ approaching zero is at the same order of the band width of the diagonal component, see the red dots in Fig.~\ref{fig:eth}. Therefore, one may conclude that ETH is satisfied, in the strong coupling limit, for both the diagonal and off-diagonal elements of the Wigner function.

\begin{figure}[!hbpt]
    \centering
    \includegraphics[width=0.23\textwidth]{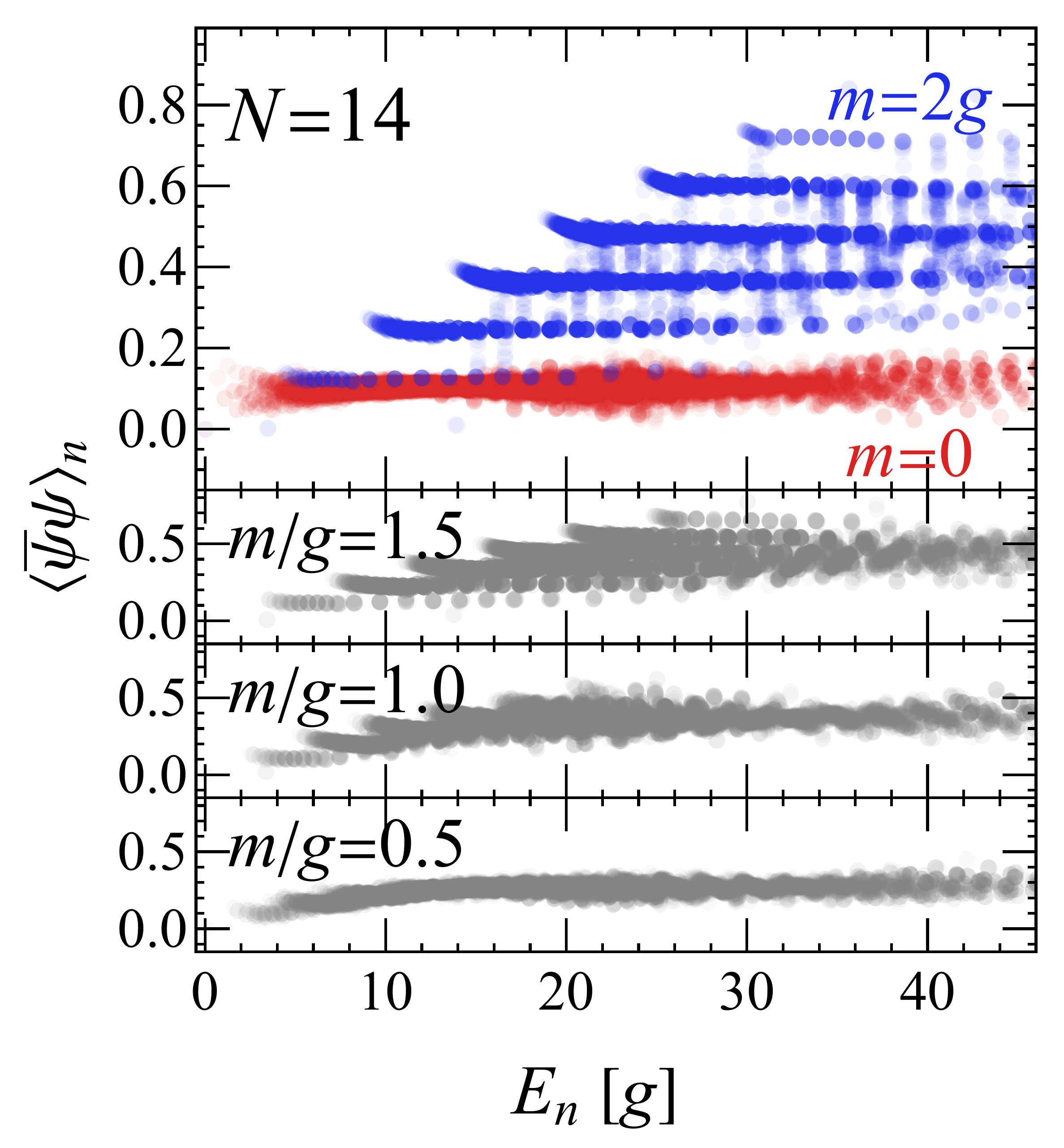}
    \includegraphics[width=0.23\textwidth]{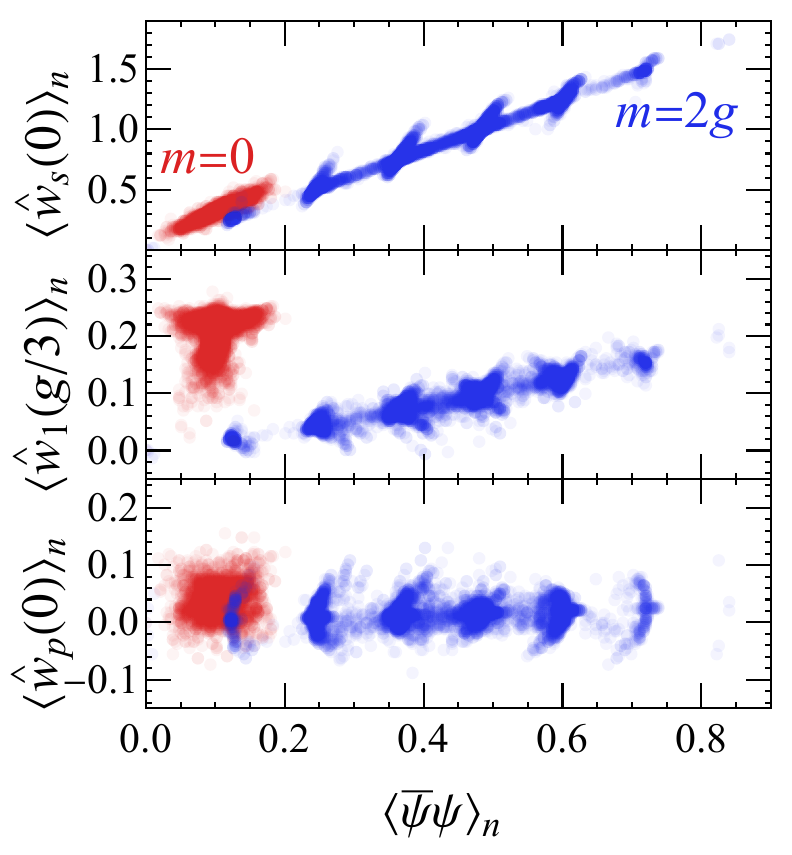}
    \caption{(Left) Same as Fig.~\protect{\ref{fig:eth2}} but for the scalar condensate.
    (Right) Correlation between scalar condensate (horizontal) and different components of the Wigner function (vertical) measured at different energy eigenstates. Again, vector components are always vanishing and not shown.
    \label{fig:eth3}}
\end{figure}
\textit{Origin of band structure.} --- 
One should have noticed the band structure in the scalar and axial components in the weak coupling cases, while such a structure is absent in the pseudoscalar and vector\footnote{Note that $\langle \hat{w}_0 \rangle_n$'s are always zero.} components. The distinction between such two categories results in the difference in thermalization --- from Figs.~\ref{fig:wigner} and ~\ref{fig:wigner_diff} one can tell that in the weak coupling case ($m=2g$), the vector and pseudoscalar Wigner functions approach to thermal equilibrium whereas the scalar and axial vector components do not.

Noting that the momentum integral of $\hat w_s(p)$ is the scalar condensate ($\bar\psi \psi$), and the mass term $m\,\bar\psi \psi$ dominates the Hamiltonian for large $m$, a natural speculation is that, in the weak coupling cases, $\bar\psi \psi$ is approximately conserved and cause the band structure in $w_s$ and $w_a$. To check this, we measure $\bar\psi \psi$ for each energy eigenstate and present the $E_n$-versus-$\langle \bar\psi \psi\rangle_n$ scatter plot in the left panel of Fig.~\ref{fig:eth3} and the $\langle \bar\psi \psi\rangle_n$-versus-$\langle \hat{w}_i \rangle_n$ in the right panel. As expected, in the upper panel, the $\langle \bar\psi \psi\rangle_n$ values concentrate at the levels $2/N$, $4/N$, $\cdots$, $14/N$, with lattice size $N=14$, corresponding to the states with one, two, $\cdots$, six fermion-antifermion pair(s), respectively. 
In the right panel which shows the relation between $\bar\psi \psi$ and different components of the Wigner function measured for each energy eigenstate, we observe the strong correlation between the scalar condensate and the scalar Wigner function in both weak and strong coupling, due to the reason mentioned in the beginning of the present paragraph. In addition, we observe the correlation between $\langle \bar\psi \psi\rangle_n$ and $\langle \hat{w}_1 \rangle_n$ but not for $\langle \hat{w}_p \rangle_n$. This is originated from the similarity in $\hat{w}_1$ and $\hat{w}_s$ operators when expressed in momentum eigenstate operators (c.f. Eq.~\ref{eq:wigner_momentum_operator}), the correlation in $\langle \hat{w}_1 \rangle_n$ leads to its band structure in Fig.~\ref{fig:eth3} and non-thermalization. 

\begin{figure}[!hbpt]
    \centering
    \includegraphics[width=0.49\textwidth]{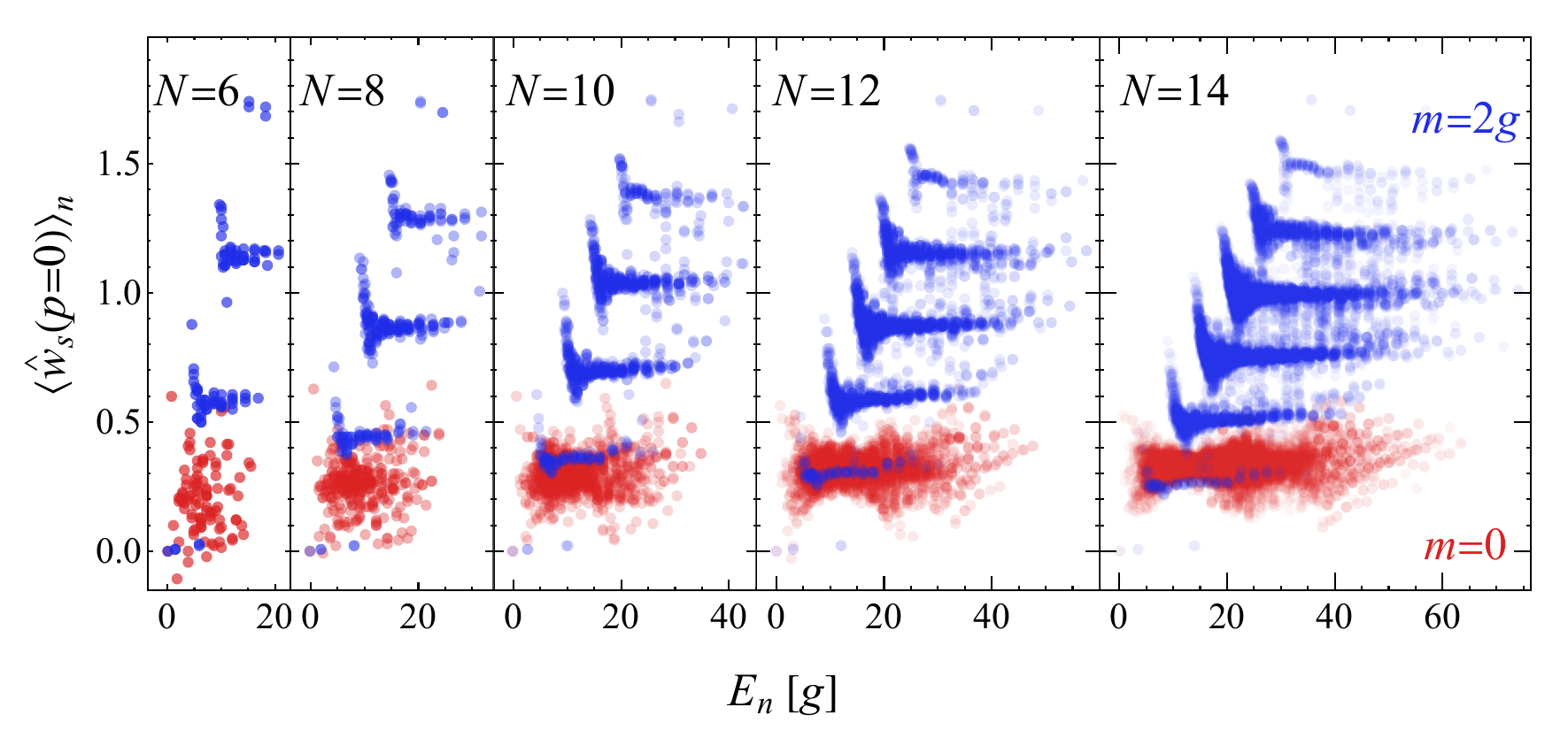}
    \caption{Same as Fig.~\protect{\ref{fig:eth}}(upper-left) but for varies lattice sizes.
    \label{fig:eth4}}
\end{figure}
\textit{Approach to continuum limit.} --- 
Finally we check the size dependence of the ETH behavior. We computed the same quantities in Fig.~\protect{\ref{fig:eth}}(upper-left) using the same parameter setting except for varies lattice sizes, and present them in Fig.~\ref{fig:eth4}. It is evident that the red dots ($m=0$) are consistent with to a smooth, narrow band, and the band width decreases when increasing $N$. In contrast, the blue dots ($m=2g$) do not exhibit narrowing in width --- when increasing $N$, while the bands are closer to each other, the total width does not decrease. One may extrapolate that in the $N\to\infty$ limit, the whole region between $0$ and $\sim 1.5$ would be filled by blue dots. Thus, we draw the conclusion that $\hat{w}_s$ satisfies ETH in the strong coupling limit but not the weak coupling $m=2g$ case. 

\section{Initial state preparation}\label{sec:app:initial}
In this Appendix, we show details in the preparation of initial state for the time evolution.
\begin{table}[!ht]
\centering
\caption{Iteration Procedure}
\begin{tabular}{cp{0.7\linewidth}}
\hline
\hline
{I.}& Randomly sample an initial condition for the coefficients $c_n$, set maximum iterate steps $n_{iter}=2000$. \\
{II.}& In each iteration, we\\
& \textit{a.} Calculate loss function $\mathcal L$;\\
& \textit{b.} Calculate gradient $\nabla_{c_n}\mathcal L$;\\
& \textit{c.} Update coefficients per \eqref{eq:update};\\
& \textit{d.} Normalize the updated coefficients;\\
& \textit{e.} Let $n_{iter}$ decrease by unity;\\
& \textit{f.} If $\mathcal L<10^{-5}$ or $n_{iter}<0$, go to {III}, otherwise back to \textit{a}. \\
{III.} & Output coefficients\\
\hline
\label{tab:iteration_procedure}
\end{tabular}
\end{table}
\begin{figure}[!h]
    \centering
    \includegraphics[width=0.4\textwidth]{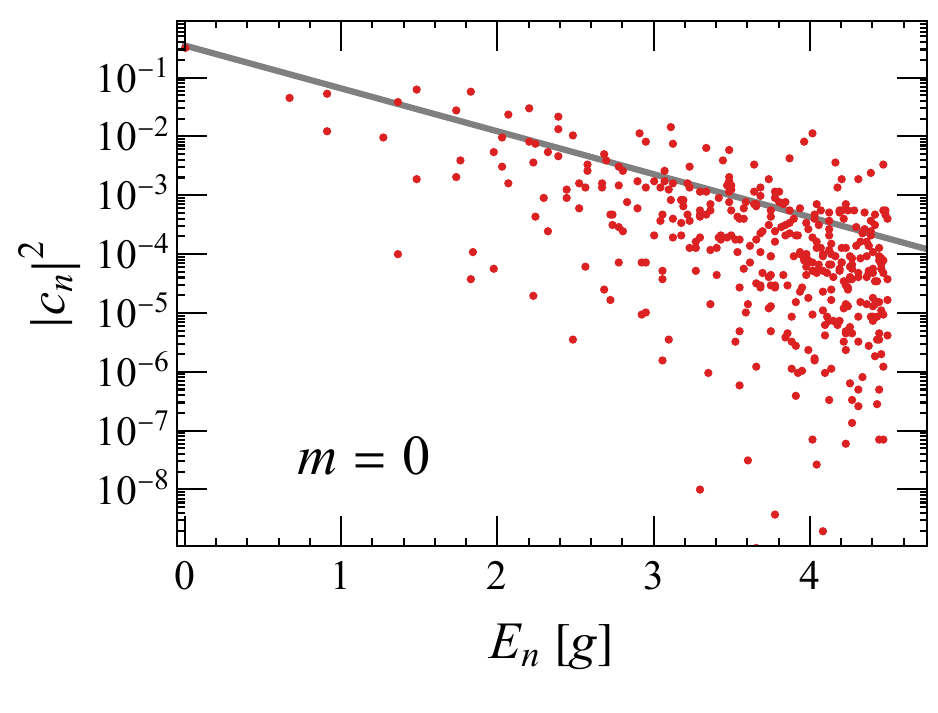}
    \caption{Probabilities of state occupation versus energy. Gray line indicates the thermal equilibrium distribution $\propto e^{- E_n / T_\mathrm{eff}}$ with the effective temperature $T_\mathrm{eff}=0.6 g$ given by reproducing the total energy.
    \label{fig:coefficient}}
\end{figure}

We begin by expanding the initial state in the form of superposition of energy eigenstates $\ket{\Psi(0)}= \sum_{n} c_n \ket{E_n}$, where $\ket{E_n}$'s are energy eigenstates and $\sum_n|c_n|^2 = 1$. Since we want certain initial expectation values for the Wigner functions so that they are as far from equilibrium as possible, we use the \textit{gradient decent} to iterate these coefficients.
The loss function is defined as
\begin{align}
\mathcal L(\{c_n\}) =\;& \sum_p\Big( |\langle\hat{w}_s(p)\rangle_0 - f_s(p)|^2 +|\langle\hat{w}_0(p)\rangle_0 - f_0(p)|^2 \nonumber\\
&\quad+|\langle\hat{w}_1(p)\rangle_0 - f_1(p)|^2 +|\langle\hat{w}_p(p)\rangle_0 - f_p(p)|^2\Big) \nonumber\\
+\;& \lambda_E \sum_n|c_n|^2 (E_n-E_0)^2\,,
\end{align}
where $\langle\hat{w}_i(p)\rangle_0 \equiv \bra{\Psi(0)}\hat{w}_i(p)\ket{\Psi(0)}$ with $i=s,p,0,1$ and $f_i$ expectation values of scalar, vector, axialvector and pseudoscalar we are setting respectively. The last term is a regularization term with energy weight to locate the least energetic state that satisfies the desired initial condition.
Then we can iterate the initial state coefficients with learning rate, $l_r=0.0001$
\begin{align}
{\bf c}^{(k+1)} =\;&{\bf c}^{(k)} - l_r {\boldsymbol{\nabla}}_{{\bf c}} \mathcal L\,,
\label{eq:update}
\end{align}
and Table~\ref{tab:iteration_procedure} lists the exact procedure we use.
In the final state, we obtained the coefficients shown in Fig.~\ref{fig:coefficient} for $m=0$, with the corresponding thermal distribution $\propto e^{-E/T_\mathrm{eff}}$ shown as the gray line. It is clear that the coefficients are far from thermal equilibrium.
\end{appendix}
\clearpage

\bibliography{ref}

\end{document}